\documentclass[]{aastex/aastex631}
\let\tablenum\relax
\usepackage{siunitx}

\newcommand\revise[1]{{#1}}

\hypersetup{
setpagesize=false,
 colorlinks=true,
 linkcolor=black,
 citecolor=black,
 filecolor=black,
 urlcolor=black,
}
\usepackage{ascmac,amsmath,amssymb,mathtools,siunitx}
\usepackage{bm}
\usepackage{comment}
\usepackage[version=3]{mhchem}
\usepackage{array,float,textcomp,ulem}
\usepackage{threeparttable}
\usepackage{cleveref}

\renewcommand{\figurename}{Fig.}
\renewcommand{\tablename}{Table~}

\newcommand{\Equref}[1]{Eq.~(\ref{#1})}
\newcommand{\Figref}[1]{Fig.~\ref{#1}}

\newcommand{\Tabref}[1]{Table~\ref{#1}}

\crefname{figure}{Fig.}{Figs.}
\crefname{equation}{Eq.}{Eqs.}
\crefname{section}{Sect.}{Sect.}

\makeatother

\def\degree{^{\circ}}

\def\beq{\begin{equation}}
\def\eeq{\end{equation}}

\def\p{\partial }
\def\d{{\rm d}}

\accepted{February 10, 2022}

\begin{document}

\title{Superrotation of Titan's stratosphere driven by the radiative heating of the haze layer}








\author{Motoki Sumi}
\affiliation{Department of Earth and Planetary Sciences, Tokyo Institute of Technology, Japan}
\altaffiliation{Current affiliation: Weathernews Inc., Japan}

\author{Shin-ichi Takehiro}
\affiliation{Research Institute for Mathematical Sciences, Kyoto University, Japan}

\author{Wataru Ohfuchi}
\affiliation{Butsuryo College of Osaka, Japan}
\affiliation{Center for Planetary Science, Kobe University, Japan}

\author{Hideko Nomura}
\affiliation{National Astronomical Observatory of Japan, Japan}

\author{Yuka Fujii}
\affiliation{National Astronomical Observatory of Japan, Japan}


\begin{abstract}

Titan's stratosphere has been observed in a superrotation state, where the atmosphere rotates many times faster than the surface does. 
Another characteristics of Titan's atmosphere is the presence of thick haze layer. 
In this paper, we performed numerical experiments using a General Circulation Model (GCM),  to explore the effects of the haze layer on the stratospheric superrotation. 
We employed a semi-gray radiation model of Titan's atmosphere following \citet[]{McKay1999}, which takes account of the sunlight absorption by haze particles.
The phase change of methane or the seasonal changes were not taken into account. 
Our model with the radiation parameters tuned for Titan 
yielded the global eastward wind around the equator with larger velocities at higher altitudes
except at around 70 km
after $10^5$ Earth days.
Although the atmosphere is not in an equilibrium state, the zonal wind profiles is approximately consistent with the observed one. 
By changing the parameters of the radiation model, we found that the intensity and the location of the maximum zonal wind velocity highly depended on the optical thickness and the altitude of the haze layer, respectively. 
Analysis on our experiments suggests that the quasi-stationary stratospheric superrotation is maintained by the balance between the meridional circulation decoupled from the surface, and the eddies that transport angular momentum equatorward.
This is different from, but similar to, the so-called Gierasch mechanism, in which momentum is supplied from the surface.
This structure may explain the no-wind region at about $80$ km in altitude. 

\end{abstract}



\section{Introduction}
\label{sec:intro}

Titan's stratosphere has been observed in a superrotation state, where the atmosphere rotates many times faster than the surface does. 
A rigorous definition of the superrotation state is given by
the index $s = m/(\Omega^2 a^2)-1$, where $a$, $\Omega$, and $m$
are, respectively, the planetary radius, rotation rate
and absolute angular momentum,  meaning the excess of absolute angular momentum 
over the equatorial value of planetary angular momentum \citep{Read1986}. 
The stratospheric superrotation of Titan's atmosphere is
prominent since the value of $s$ becomes $\sim 10$ in the equatorial region
while it is up to $\sim 0.04$ for the Earth's atmosphere \citep{Read+2018,Imamura+2020}. 
The Huygens Doppler Wind Experiment (DWE) probed Titan's atmosphere near the equator and observed the eastward winds at all altitudes up to 150~km \citep[]{Bird2005}. 
The wind velocity tends to increase with the altitude, reaching 100 m/s at around 120 km in altitude, except for the region between 60 and 100 km altitude where the wind speed is  close to zero
which has been called the ``zonal wind collapse'' region \citep[]{li2012}.
The zonal wind in the even higher altitudes ($>$ 120 km) has been constrained by the temperature measurement by Cassini Infrared Spectrometer (CIRS) with the gradient wind equation \citep[]{Achterberg2008}. 
It is found that the zonal wind velocity continues to increase up to 0.1~mbar
($\sim 300$ km)
where the wind velocity is larger than 190 m/s. 
The combination of these two observations indicates that the Titan's atmosphere is in the superrotation state from the surface through at least $\sim $ 600 km where the zonal wind 
speed increases with the altitude except the ``zonal wind collapse'' region mentioned above.

Generation and maintenance of superrotation above the equator has been regarded as one of the mysteries of atmospheric dynamics, because the friction between the atmosphere and the surface always acts to decelerate the superrotation. 
Thus, superrotation in steady state requires some mechanism that
counteracts the deceleration by transporting positive angular momentum from the surface to the atmosphere. 
\begin{figure}
    \centering
    \includegraphics[width=\linewidth]{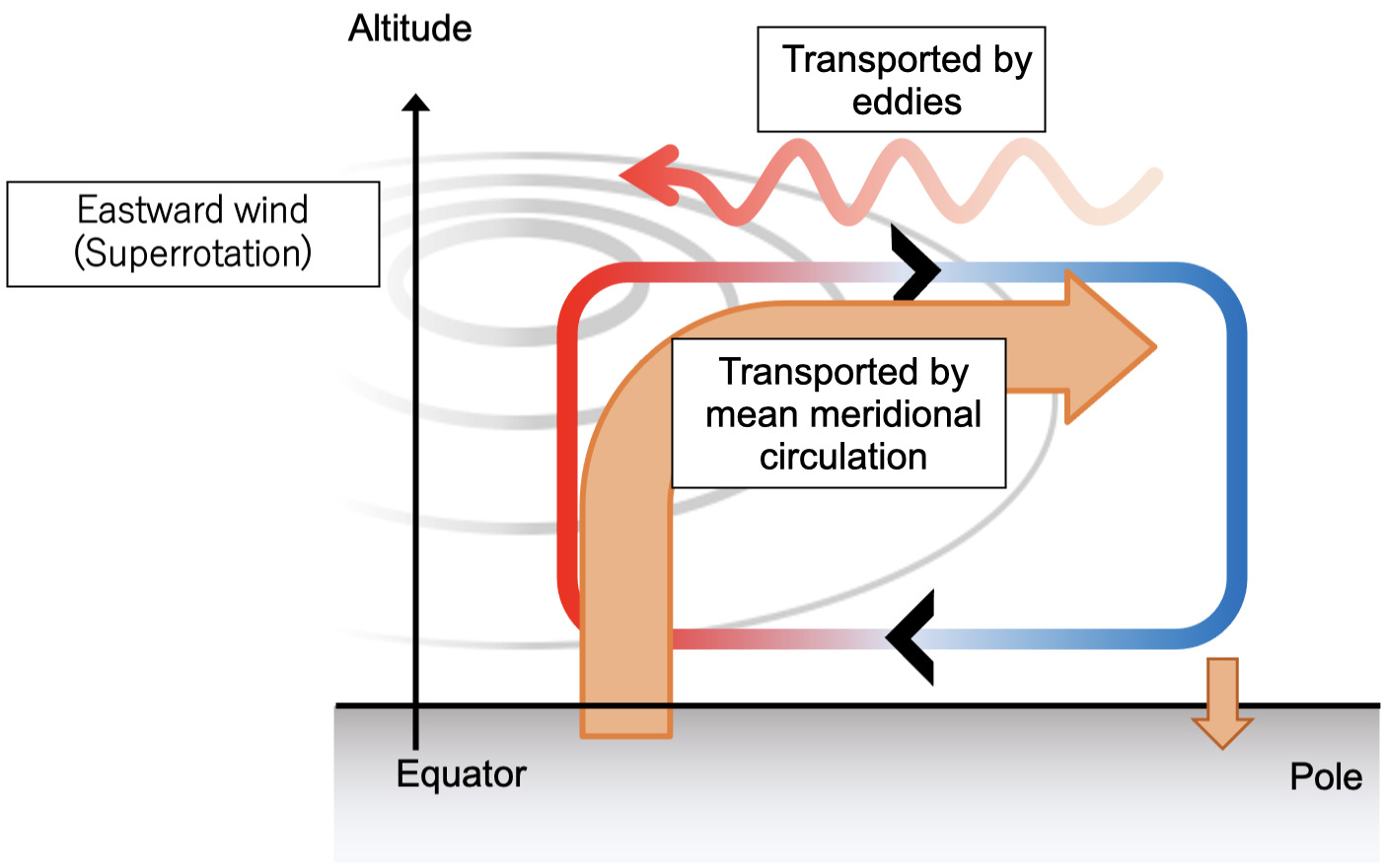}
    \caption{Schematic diagram of the Gierasch mechanism. Angular momentum transported from the surface to the upper polar side by meridional circulation is returned to the upper equator by horizontal diffusion. Eastward angular momentum is dominant in the entire atmosphere.}
    \label{fig:Gierasch}
\end{figure}
One of such scenarios proposed so far is the Gierasch mechanism \cite[]{Gierasch1975} (\Figref{fig:Gierasch}).
The key assumptions of this scenario are the meridional circulation symmetric about the equator and 
efficient horizontal diffusion of angular velocty compared to the vertical diffusion, thermal diffusion, and angular momentum advection by meridional circulation.
The first step of this scenario is that the meridional circulation transports the angular momentum from the surface to the upper atmosphere in the equatorial region.
The conservation of the {\it angular momentum} means that as the air moves along the upper layer to the polar region the {\it angular velocity} increases.
Then, the assumed efficient horizontal diffusion of angular velocity kicks in and tries to make the {\it angular velocity} of the upper atmosphere uniform by transporting the angular momentum from the polar region to the equatorial region. 
In real atmospheres, the efficient horizontal diffusion is not likely to be valid; Instead, some kind of turbulence in the atmosphere may act as the horizontal diffusion to transport the angular momentum. 

In order to understand the dynamical processes that are taking place in Titan's atmosphere, several General Circulation Models (GCMs) have been developed, including Titan WRF Model \citep[e.g.,][]{Newman2011}, IPSL Titan GCM  \citep[e.g.,][]{Lebonnois2012}, and Titan Atmosphere Model (TAM) \citep[][]{Lora2015}.
The model configuration and the elemental processes considered depend on the model used as well as purpose of each study.
These numerical experiments have successfully reproduced the approximate profile of superrotation and the meridional circulation.
%
The interpretation of the results is complicated by Titan's seasonal cycle\footnote{Titan has Earth-like obliquity with respect to Saturn's orbital plane.} in these simulations.
In Titan's summer and winter the meridional circulation is dominated by the pole-to-pole circulation (i.e., one cell), while at the equinoxes the meridional circulation is more symmetric about the equator with two cells where the air rises around the equator and sinks near the poles.
The orbital average of the simulated structures, however, indicated that the net meridional circulation transports the angular momentum from lower atmospheres to the stratosphere at low latitudes and then to the polar regions.
Thus, those studies suggest that the superrotation of Titan is maintained by the Gierasch mechanism associated with the global meridional circulation.

Nevertheless, a closer comparison between the observed profile of Titan's superrotation and the numerical results shows minor but non-negligible differences. 
The most significant difference appears in the ``zonal wind collapse'' region, i.e., the region between 60 km and 100 km.
Such a feature is absent in the results of \citet[]{Newman2011} and  \citet[]{Lora2015}, and marginally appears with much smaller amplitude in the result of \citet[]{Lebonnois2012}.

These models included multiple processes altogether, which prevents us from examining the individual factors that control the zonal wind profile.
In other words, it is not clear what are the key properties or processes to generate and maintain superrotation. 
This motivates a GCM study with a simpler configuration to explore  the effects of the individual properties/processes of Titan's atmosphere on the superrotation. 

Studies in this direction include \cite{MitchellVallis2010}, 
which investigated the dependence of atmospheric circulation on the thermal Rossby number, using Newtonian cooling model where the forcing profile simply consists of an adiabatic troposphere and an isothermal stratosphere. 
The authors found that strong superrotation develops in the troposphere at Titan-like large thermal Rossby numbers and  is maintained by the angular momentum transport by global barotropic waves. 
Using a similar model, \cite{Mitchell+2014} found that the seasonal cycle acts to prevent superrotation for Titan-like parameters. 
While these simulations provide useful insights into the general processes that govern the development of superrotation with varying atmospheric parameters, the assumed thermal forcing can be very different from that in Titan's atmosphere, and the effect of Titan's unique thermal profile on the superrotation state remains unclear. 

The goal of this study is to further develop our understanding of the elemental processes that contribute to Titan's superrotation using a model that is relatively simple but includes a Titan-like thermal profile.
In particular, we focus on the effects of the haze layer on the generation and maintenance of superrotation. 
For this purpose, we perform numerical experiments using a General Circulation Model (GCM) by introducing 
a semi-gray radiation model of Titan following  \citet[]{McKay1999}, which takes account of the sunlight absorption by the haze layer. 
The phase change of methane and the seasonal cycle are not taken into account so that the effects of the radiation field structure induced by the haze layer are isolated for examination.
Moreover, in order to clarify the relation between the haze layer and superrotation, we vary the parameters related to the haze layer and analyze the effects on superrotation.
This aspect is a contrast to the previous studies which fixed the properties of the haze layer. 
Specifically, we varied the solar absorption efficiency in the haze layer and the altitude at which the haze exists, and compared the behavior of the atmospheric dynamics. 

The organization of the paper is as follows.
In section 2, we introduce the model and settings of our calculations. 
In section 3, we show the results of our calculations, and analyze the results and investigate the structure of superrotation maintenance. 
Finally, in section 4, we 
discuss our results in comparison with the related numerical simulations of the Venusian atmosphere and observed structure of Titan's atmosphere.


\section{Method}
\label{s:method}

In this section, we describe the computational model used in this study and the parameters assumed. 

\subsection{Dynamical core and model grids}
\label{sec:dynamics}

We used the planetary atmospheric general circulation model DCPAM5 \citep{dcpam} to calculate the planetary-scale atmospheric general circulation. 
DCPAM5 adopts the primitive equations on a rotating sphere with the $\sigma $ coordinate system \citep{2006igsm.book.....K}. 
The set of equations used in DCPAM5 is summarized in Appendix \ref{ap:dcpam5}. 


The horizontal resolution of our standard model is 32 longitudinal grid points and 16 latitudinal grid points (the spectral truncation of so-called T10).
We additionally ran the model with a higher resolution (64 $\times $ 32; T21) to check the resolution dependence.

The vertical grid of the model is expressed in terms of $\sigma \equiv p/p_{\rm surf}$, the ratio of the pressure $p$ to the surface pressure $p_{\rm surf}$, and contains 55 levels. 
The intervals of $\sigma$ levels are equal with respect to $p$ in the lower atmosphere ($\sigma > 0.1 $) while they are equal with respect to $\log p$ in the upper atmosphere ($\sigma < 0.1$). 
In our runs, the vertical grid points covers the altitude from the surface to a height of approximately 400 km.
As an example, 
\Tabref{tb:vertical layer} shows the altitudes of the grid points calculated based on one of the experiments of our study. 

\begin{deluxetable*}{cccccccc}
\tablecaption{Sigma, and globally averaged pressure and altitude of an experiment at the 55 vertical layers of the GCM.\label{tb:vertical layer}}
\tablewidth{0pt}
\tabletypesize{\scriptsize}
\tablehead{
\colhead{Layer} &
\colhead{$\sigma$} &
\colhead{Pressure} &
\colhead{Altitude} &
\colhead{Layer} &
\colhead{$\sigma$} &
\colhead{Pressure} &
\colhead{Altitude} \\
\colhead{} &
\colhead{} &
\colhead{(Pa)} &
\colhead{(km)} &
\colhead{} &
\colhead{} &
\colhead{(Pa)} &
\colhead{(km)} 
}
\startdata
1 & 9.827e-1 & 1.442e+5 & 0.3581 & 28 & 5.877e-2 & 8.622e+3 & 48.46 \\ 
2 & 9.482e-1 & 1.391e+5 & 1.095 & 29 & 4.173e-2 & 6.122e+3 & 53.86 \\ 
3 & 9.138e-1 & 1.341e+5 & 1.871 & 30 & 2.963e-2 & 4.347e+3 & 60.00 \\ 
4 & 8.793e-1 & 1.290e+5 & 2.669 & 31 & 2.104e-2 & 3.087e+3 & 67.52 \\ 
5 & 8.448e-1 & 1.239e+5 & 3.466 & 32 & 1.494e-2 & 2.192e+3 & 76.48 \\ 
6 & 8.103e-1 & 1.189e+5 & 4.324 & 33 & 1.061e-2 & 1.556e+3 & 87.26 \\ 
7 & 7.758e-1 & 1.138e+5 & 5.148 & 34 & 7.534e-3 & 1.105e+3 & 98.62 \\ 
8 & 7.413e-1 & 1.087e+5 & 6.107 & 35 & 5.349e-3 & 7.847e+2 & 110.8 \\ 
9 & 7.068e-1 & 1.037e+5 & 7.007 & 36 & 3.799e-3 & 5.573e+2 & 123.5 \\ 
10 & 6.724e-1 & 9.864e+4 & 7.964 & 37 & 2.697e-3 & 3.956e+2 & 136.0 \\ 
11 & 6.379e-1 & 9.358e+4 & 8.965 & 38 & 1.915e-3 & 2.809e+2 & 148.7 \\ 
12 & 6.034e-1 & 8.852e+4 & 10.02 & 39 & 1.360e-3 & 1.995e+2 & 162.0 \\ 
13 & 5.689e-1 & 8.346e+4 & 11.17 & 40 & 9.657e-4 & 1.417e+2 & 175.3 \\ 
14 & 5.344e-1 & 7.840e+4 & 12.27 & 41 & 6.857e-4 & 1.006e+2 & 188.3 \\ 
15 & 4.999e-1 & 7.334e+4 & 13.44 & 42 & 4.869e-4 & 7.143e+1 & 201.4 \\ 
16 & 4.654e-1 & 6.827e+4 & 14.77 & 43 & 3.457e-4 & 5.071e+1 & 214.8 \\ 
17 & 4.310e-1 & 6.323e+4 & 16.20 & 44 & 2.455e-4 & 3.601e+1 & 227.7 \\ 
18 & 3.965e-1 & 5.817e+4 & 17.53 & 45 & 1.743e-4 & 2.557e+1 & 241.7 \\ 
19 & 3.620e-1 & 5.311e+4 & 19.29 & 46 & 1.238e-4 & 1.816e+1 & 254.6 \\ 
20 & 3.275e-1 & 4.804e+4 & 20.87 & 47 & 8.790e-5 & 1.289e+1 & 267.5 \\ 
21 & 2.930e-1 & 4.298e+4 & 22.66 & 48 & 6.241e-5 & 9.156e+0 & 281.5 \\ 
22 & 2.585e-1 & 3.792e+4 & 24.82 & 49 & 4.432e-5 & 6.502e+0 & 295.9 \\ 
23 & 2.240e-1 & 3.286e+4 & 27.18 & 50 & 3.147e-5 & 4.617e+0 & 311.0 \\ 
24 & 1.895e-1 & 2.780e+4 & 30.01 & 51 & 2.235e-5 & 3.279e+0 & 325.5 \\ 
25 & 1.550e-1 & 2.274e+4 & 33.04 & 52 & 1.587e-5 & 2.328e+0 & 340.1 \\ 
26 & 1.204e-1 & 1.766e+4 & 37.07 & 53 & 1.127e-5 & 1.653e+0 & 354.4 \\ 
27 & 8.580e-2 & 1.259e+4 & 42.47 & 54 & 8.000e-6 & 1.174e+0 & 369.5 \\ 
\nodata & \nodata & \nodata & \nodata & 55 & 2.773e-6 & 4.068e-1 & 399.4 \\ 
\enddata
\end{deluxetable*}


\subsection{Radiative transfer}
\label{sec:heating}


The primitive equations (Appendix \ref{ap:dcpam5}) require the heating rate as input, which is calculated as follows. 
We consider radiative transfer in the shortwave band (i.e., the wavelengths where the solar flux dominates over the thermal emission) and that in the longwave band (the wavelengths where the thermal emission dominates) separately.
The wavelength dependence in each band is, however, largely ignored for simplicity; In this sense, our model can be regarded as a ``semi-gray'' model.
Our model ignores scattering within the atmosphere and by the surface, and the only Bond albedo at the top of the atmosphere is taken into account.
The longwave and shortwave equations are described in the following.


For a non-scattering atmosphere, the radiative transfer equations for the upward and downward longwave fluxes as function of longwave optical depth, $F_L^{+/-} (\tau _L)$, are
\begin{align}
    F_L^+(\tau_L) &= \pi B (\tau_{{\rm surf},L}) \mathcal{T} (\tau_{{\rm surf},L}, \tau) - \int^{\tau_{{\rm surf},L}}_{\tau_L} \pi B(\tau')\dfrac{d\mathcal{T}(\tau_L, \tau')}{d\tau'}d\tau', \\
    F_L^-(\tau_L) &= \int^{\tau_L}_0 \pi B(\tau')\dfrac{d\mathcal{T}(\tau_L, \tau')}{d\tau'}d\tau',
\end{align}
where $\tau _{{\rm surf},L}$ is the optical depth at the surface and \(\mathcal{T}(\tau_L, \tau') \equiv \exp{(-3/2\,(\tau_L - \tau'))}\) is the transmission coefficient.
The $B (\tau )$ is the band-integrated Planck function for the temperature at $\tau $, which is simply 
\begin{align}
    \pi B (\tau) &= \sigma_{\rm SB}T^4(\tau),
\end{align}
where \(\sigma_{\rm SB}\) is the Stefan-Boltzmann constant, and $T$ is temperature.

The longwave optical depth is related with the pressure by a power law, based on the previous studies with gray atmospheres \citep[e.g.,][]{Pollack+1969, McKay1999}:
\begin{align}
    \tau_{L} (p) = \tau_{{\rm surf},L} \left( \dfrac{p}{p_{\rm surf}}\right)^n. \label{eq:tau_L}
\end{align}
Note that because the haze layer is optically thin in the longwave, the longwave opacity is simply due to the absorption by atmospheric molecules, consistent with the previous studies that adopts \Equref{eq:tau_L}.


The shortwave radiative transfer is where the direct effect of the haze layer comes in.
The haze layer is a significant absorber of the sunlight, altering the temperature structure of the upper atmosphere of Titan.
%
The first parameter to represent the effect of the haze layer is the shortwave optical depth, $\tau _S$, which is largely due to the opacity of the haze layer.
Following \citet{McKay1999}, the shortwave optical depth of Titan is approximately related to the longwave optical depth by
\begin{equation}
    \tau _S (\tau_L) = k \tau_L.
\end{equation}
Furthermore, an additional parameter is introduced to control the fraction of the sunlight that is affected by the haze layer, $\gamma $ ($0 \leq \gamma \leq 1$).
Physically, this may be roughly interpreted as the fraction of the wavelengths for absorption by the haze layer, or the haze cover fraction.
Adopting these parameters, the shortwave radiative transfer equations for a non-scattering atmosphere are
\begin{align}
    F_S^+ &= 0. \\
    F_S^- (\tau _L) &= \tilde F_{TOA} \, \gamma \,  e^{-\tau_S(\tau_L)} + \tilde F_{TOA} \, (1-\gamma) \,,
\end{align}
where \(\tilde F_{TOA}\) is the solar flux on the top of the atmosphere multiplied by $( 1 - A )$ with \(A\) being the Bond albedo.
Here, we assume that the surface albedo is zero. 
Titan’s surface albedo is estimated to be about 0.1-0.15 by Huygens observations \citep{2016Icar..270..260K}, and \citet{Lebonnois2012} assumed 0.15 in their GCM simulations. 
When the albedo is increased from zero, the circulation in the troposphere would be reduced; however, it would not affect the upper atmosphere that we mainly pay attention to, since the radiative effect of the haze layer is dominant there. 

Once the vertical structure of the flux is determined, the heating rate due to the radiation process is obtained by
\begin{align}
    Q &= - \dfrac{1}{C_p \rho} \dfrac{\partial F}{\partial z} = \dfrac{g}{C_p} \dfrac{\partial F}{\partial p} \label{eq:Heating rate}, \\
    F &= ( F_L^{+} - F_L^{-} ) + ( F_S^{+} - F_S^{-} ),
\end{align}
This is the input of \Equref{eq:thermodynamic}.


\subsection{Input parameters and assumptions}
\label{sec:setup}

\subsubsection{Constants}

\Tabref{tb:constants_1} summarizes the input values for the planetary parameters that are kept constant throughout the experiments.

\begin{deluxetable*}{ccc}[htbp]
\tabletypesize{\scriptsize}
\tablecaption{Planetary parameters}
\tablewidth{0pt}
\label{tb:constants_1}
\tablehead{
\colhead{Description} &
\colhead{Symbol} &
\colhead{Value (units)}
}
\startdata
Radius of planet 	& $a$ & $2.575 \times 10^6 $ \ (${\rm  m }$)\\
Gravitational acceleration & $g$ & 1.35 \ (${\rm m \ s^{-2}}$)\\
Angular velocity & $\Omega$ & $4.57329 \times 10^{-6}$ \ (${\rm s^{-1}}$)\\
Gas constant of air & $R$ & 8.31 \ (${\rm  J \ K^{-1} \ mol^{-1} }$) \\
Specific heat of air at constant pressure   & $C_p$ & 1040.0 \ (${\rm J \ kg^{-1} \ K^{-1}}$)\\
Mean molucular weight of air & $M$ & $27.3 \times 10^{-3}$ \ (${\rm kg \ mol^{-1}}$) \\
Pressure at surface & $p_{\rm surf}$ & $1.467 \times 10^5$ \ (${\rm Pa}$)\\
Optical depth at surface & $\tau_{\rm surf}$ & 3.0 \\
Bond Albedo & $A$ & 0.3 \\
Solar Constant & $F_{\rm sun}$ & 14.0 \ (${\rm W \ m^{-2}}$) \\
\enddata
\end{deluxetable*}


\subsubsection{Radiative transfer parameters that are varied}

Our radiative transfer equations have a few parameters to be prescribed: $n$, $k$, and $\gamma$.
To find the standard values tuned for Titan's atmosphere, we computed 1-dimensional radiative-equilibrium temperature profiles by varying these three parameters searching for the set of values that reasonably reproduces the observed data.
As a result, we set $(n, k, \gamma )=(1.4, 140, 0.44)$ as the values for Titan's atmospheres. 

\begin{figure}[htbp]
    \centering
    \includegraphics[width=0.5\linewidth]{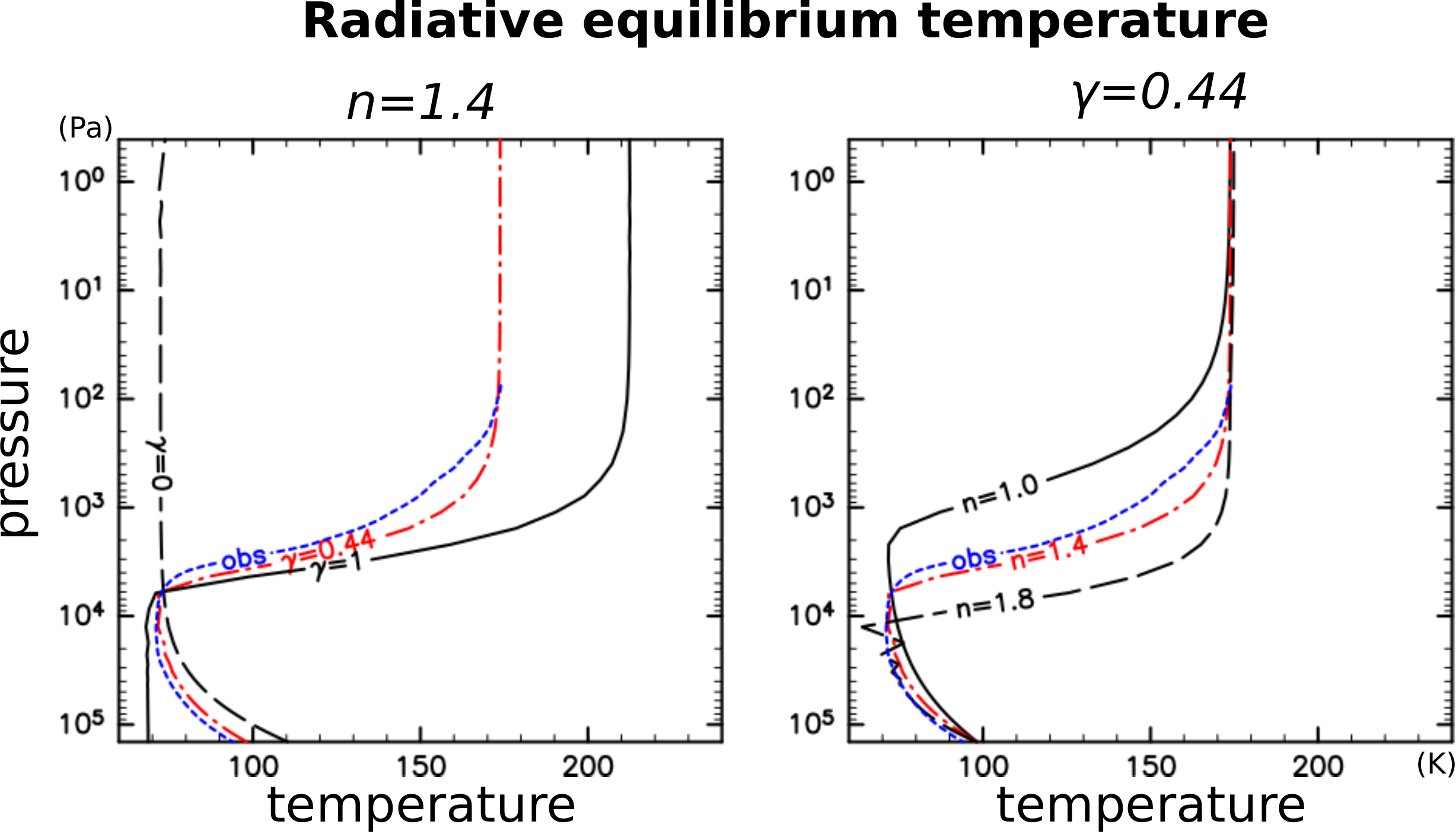}
    \caption{The equilibrium temperature profiles for $\gamma, n$, where Left panel is the profile for $\gamma$ ($n$ is fixed), Right panel is the profile for $n$ ($\gamma$ is fixed).
    Blue lines show the temperature distribution from the observation \citep{Lellouch+1989}.
    }
    \label{fig:initial_temperature}
\end{figure}

In addition, we change $n$ and $\gamma $ around these fiducial values as summarized in \Tabref{tb:gamma} and \Tabref{tb:n} in order to explore the effect of different profiles of the haze layer.
The corresponding radiative equilibrium solutions are shown in \Figref{fig:initial_temperature}.
As the figure indicates, $\gamma $ controls the amount of the solar flux absorbed by the haze layer, while varying $n$ essentially changes the altitudes that are heated up due to the haze layer.


\begin{deluxetable*}{ccc}[htbp]
\tabletypesize{\scriptsize}
\caption{Parameter $\gamma$ for the solar absorption rate in the haze layer. In this case, $n=1.4$ is fixed.\label{tb:gamma}}
\tablewidth{0pt}
\label{tb:constants_3}
\tablehead{
\colhead{Description} &
\colhead{$\gamma$}
}
\startdata
No haze layer & 0  \\
Titan haze layer & 0.44 \\
Only haze layer & 1 \\
\enddata
\end{deluxetable*}

\begin{deluxetable*}{ccc}[htbp]
\tabletypesize{\scriptsize}
\caption{Parameter $n$ for the altitude of the haze layer. In this case, $\gamma=0.44$ is fixed. \label{tb:n}}
\tablewidth{0pt}
\label{tb:constants_4}
\tablehead{
\colhead{Description} &
\colhead{$n$}
}
\startdata
Higher haze layer & 1.0  \\
Titan haze layer & 1.4 \\
Lower haze layer & 1.8 \\
\enddata
\end{deluxetable*}


\subsubsection{Boundary conditions}
\label{sss:boundary_TOA}

We assume that the insolation at the top of the atmosphere is longitudinally uniform.
This is a reasonable assumption because the radiative relaxation time of Titan is much longer than the rotation period (approximately 16 Earth days).
The radiative relaxation time, $\tau_r$, is expressed as,
\begin{align}
    \tau_r \sim \left( \dfrac{C_p}{R} \right) \dfrac{P\Delta H}{\sigma_{SB}T^4}, \label{eq:radiative_timescale}
\end{align}
where $P$ and $\Delta H$ are the typical values of pressure and the atmosphere thickness, respectively.
It is about $2\times10^5$ days near the surface for Titan's parameters ($C_P/R\sim 1$, $P\sim10^5$ Pa, $\Delta H \sim 10^6$ m, $T\sim100$ K).

We ignore the seasonal variation, i.e., the insolation is symmetric with respect to the equator, in order to simplify the system and focus on the effect of the haze layer. 
As mentioned in section \ref{sec:intro}, however, previous work found that Titan's seasonal cycle rather acts to prevent superrotation \citep{Mitchell+2014}. Introducing seasonal cycles to our model could therefore have non-negligible effects on atmospheric dynamics. Such effects are scopes of the future work. 

The setup related to the planetary surface is rather simplified compared to that generally used in realistic GCM simulations. 
We do not include topography. 
The atmosphere and the surface exchange energy through radiation only; The surface temperature is determined by the solar and thermal radiation reaching the surface. 
Specifically, we do not include the sensible or latent heat exchange between the atmosphere and the surface. 
While the omission of them may have a minor effect on the circulation near the surface, it is not likely to affect the structure at the upper atmospheres we focus on.

\subsubsection{Parametrization for diffusion, dissipation, and convection}
\label{sss:diff_diss_fric}

The equation system of our model is solved for the vorticity, $\zeta $, and the divergence, $D$, of the horizontal velocity field, 
with the temperature and surface pressure, $T$ and $p_s$. 
To suppress the numerical instabilities due to wave turbulence, we include the horizontal diffusion of $\zeta$, $D$, and $T$ in a usual manner. 
The horizontal diffusion terms are expressed as follows:
\begin{align}
    \mathcal{D}_{HD}(\zeta) &= -K_{HD}\left[ (-1)^{N_D/2}\nabla^{N_D}-\left( \dfrac{2}{a^2} \right)^{N_D/2} \right] \zeta, \\
    \mathcal{D}_{HD}(D) &= -K_{HD}\left[ (-1)^{N_D/2}\nabla^{N_D}-\left( \dfrac{2}{a^2} \right)^{N_D/2} \right] D, \\
    \mathcal{D}_{HD}(T) &= -K_{HD}\left[ (-1)^{N_D/2}\nabla^{N_D}\right] T, \\
    K_{HD} &= \dfrac{1}{\tau_{HD}}\left(\dfrac{n_{\rm max}(n_{\rm max}+1)}{a^2}\right)^{-N_D/2},
    \label{eq:diffusion}
\end{align}
where $a$ is the planetary radius and $\tau_{HD}$ is the timescale of horizontal diffusion for the components with the maximum total horizontal wavenumber, $n_{\rm max}$ (the resolution of our model is T10 or T21,  i.e., $n_{\rm max}=10$ or $21$). 
Larger $N_D$ makes diffusion more effective for large wavenumber components. 
We use the value of the order of the horizontal diffusion $N_D$ of 6. Thus, the horizontal diffusion of our model is also called hyper viscosity and diffusion. 
On the other hand, the dissipation in the so-called ``sponge layer'' is written as
\begin{align}
    \mathcal{D}_{SL}(\zeta) &= -\gamma_M (\zeta - \overline{\zeta}),\\
    \mathcal{D}_{SL}(D) &= -\gamma_M(D-\overline{D}),\\
    \mathcal{D}_{SL}(T) &= -\gamma_H (T - \overline{T}).
\end{align}
where $\overline{\zeta }$, $\overline{D}$ and $\overline{T}$ represent the zonal mean. 
The coefficient $\gamma _M$  and  $\gamma _H$ are assumed to decay as a power-law of $\sigma \equiv p/p_s$ from the model-top, $\sigma _0$, down to the lower boundary of the ``sponge layer'', $\sigma _{\rm lim}$:
\begin{align}
    \gamma_M= \begin{cases}
        \gamma_{M,0} \left( \dfrac{\sigma_0}{\sigma}\right)^{N_{SL}}, & (\sigma \le \sigma_{\rm lim}) \\
        0. & (\sigma > \sigma_{\rm lim})
    \end{cases}\\
    \gamma_H= \begin{cases}
        \gamma_{H,0} \left( \dfrac{\sigma_0}{\sigma}\right)^{N_{SL}}, & (\sigma \le \sigma_{\rm lim}) \\
        0. & (\sigma > \sigma_{\rm lim})
    \end{cases}
\end{align}

Furthermore, the friction between the atmosphere and the surface is included in the form of Rayleigh friction:
\begin{align}
    \mathcal{F}_\lambda &= -k_{\rm v}u, \\
    \mathcal{F}_\varphi &= -k_{\rm v}v. 
\end{align}
where $u$ and $v$ are the horizontal velocities. The friction coefficient, $k_{\rm v}$, is assumed to decrease as a linear function of $\sigma $ until it becomes zero at $\sigma _B$, i.e., 
\begin{align}
    k_{\rm v} = \begin{cases}
        \dfrac{1}{\tau_f} \dfrac{\sigma - \sigma_B}{1.0-\sigma_B}, & (\sigma > \sigma_B)  \\
        0, & (\sigma \le \sigma_B)
    \end{cases}
\end{align}
where $\tau_f$ corresponds to the time constant for the friction.

The way these terms are included in the equations of motion can be found in Appendix (\Equref{eq:motion of u}, \Equref{eq:motion of v}, \Equref{eq:U_A} and \Equref{eq:V_A}). 
The values assumed for the parameters in these terms are summarized in \Tabref{tb:constants_5}. 

An adjustment scheme is used for the parameterization of sub-grid scale dry convection \revise{based on the concept proposed by} \cite{Manabe1965}. 
It is effective especially for avoiding numerical divergence near the top of the atmosphere. 

\begin{deluxetable*}{ccc}[tbhp]
\tabletypesize{\scriptsize}
\tablecaption{Numerical parameters for diffusion, sponge layer and surface drag}
\tablewidth{0pt}
\label{tb:constants_5}
\tablehead{
\colhead{Description} &
\colhead{Symbol} &
\colhead{Value (units)}
}
\startdata
Time scale for the horizontal diffusion & $\tau_{HD}$  & 1.0 \ ($\rm day$) \\
Order of the horizontal diffusion & $N_D$ & 6 \\
Time scale for the sponge layer &  $\gamma_{M,0}$, $\gamma_{H,0}$  & 1.0 \ ($\rm day^{-1}$) \\
Order of the sponge layer & $N_{SL}$ & 1\\
Bottom limit of the sponge layer & $\sigma_{\rm lim}$  & $1.127\times 10^{-5}$ \\
Top height where the surface drag operates & $\sigma_B$ & 0.8 \\
Time scale for the surface drag & $\tau_f$ & $1.0 \times 10^2$ \ ($\rm day$) \\
\enddata
\end{deluxetable*}

\subsection{Running the model}

Due to the long radiative timescale (\Equref{eq:radiative_timescale}), running the model from the initial condition where the temperature is uniform would take a long computational time for convergence.
To reduce the computation time, we run the model with the radiative transfer calculation only (turning off the dynamical calculation) for the first $10^5$ Earth days and obtain a temperature profile in a radiative equilibrium. 
In this calculation we give the horizontally (i.e., not only longitudinally but also latitudinally) uniform top-of-atmosphere insolation ($F_{\rm sun}/4$). 
Thus, this is in practice a single column (i.e. 1D radiative-convective) calculation. 
Using this temperature structure as well as the zero wind velocity everywhere, and puting back the latitudinal gradient of  insolation, we resume the run with the dynamical calculation turned on. 
The reason for assuming the uniform insolation in the initial radiation-only calculation is because when the latitudinal gradient of insolation is included the resultant temperature gradient is so large that the following all-in calculation becomes very unstable. 

We stopped each dynamical calculation after additional $10^5$ Earth days, corresponding to 10 Titan years. 
This is approximately the radiative relaxation timescale near the surface (\Equref{eq:radiative_timescale}) and is substantially longer than the seasonal timescale of Titan. 
We analyzed the atmospheric structure in detail at this anchoring point, because as we will show later we found an interesting properties that are reminiscent of Titan's. 
However, we also found that the atmosphere did not reach an equilibrium at this point. Therefore, we also continued the calculation until full equilibration ($10^6$ days). We will briefly present the atmospheric structure in this final state as well.

\section{Results}

\subsection{Dependence on the solar absorption efficiency of haze layer}
\label{ss:dependence_gamma}

\subsubsection{Atmospheric structure}

\begin{figure}[htbp]
    \centering
    \includegraphics[width=\linewidth]{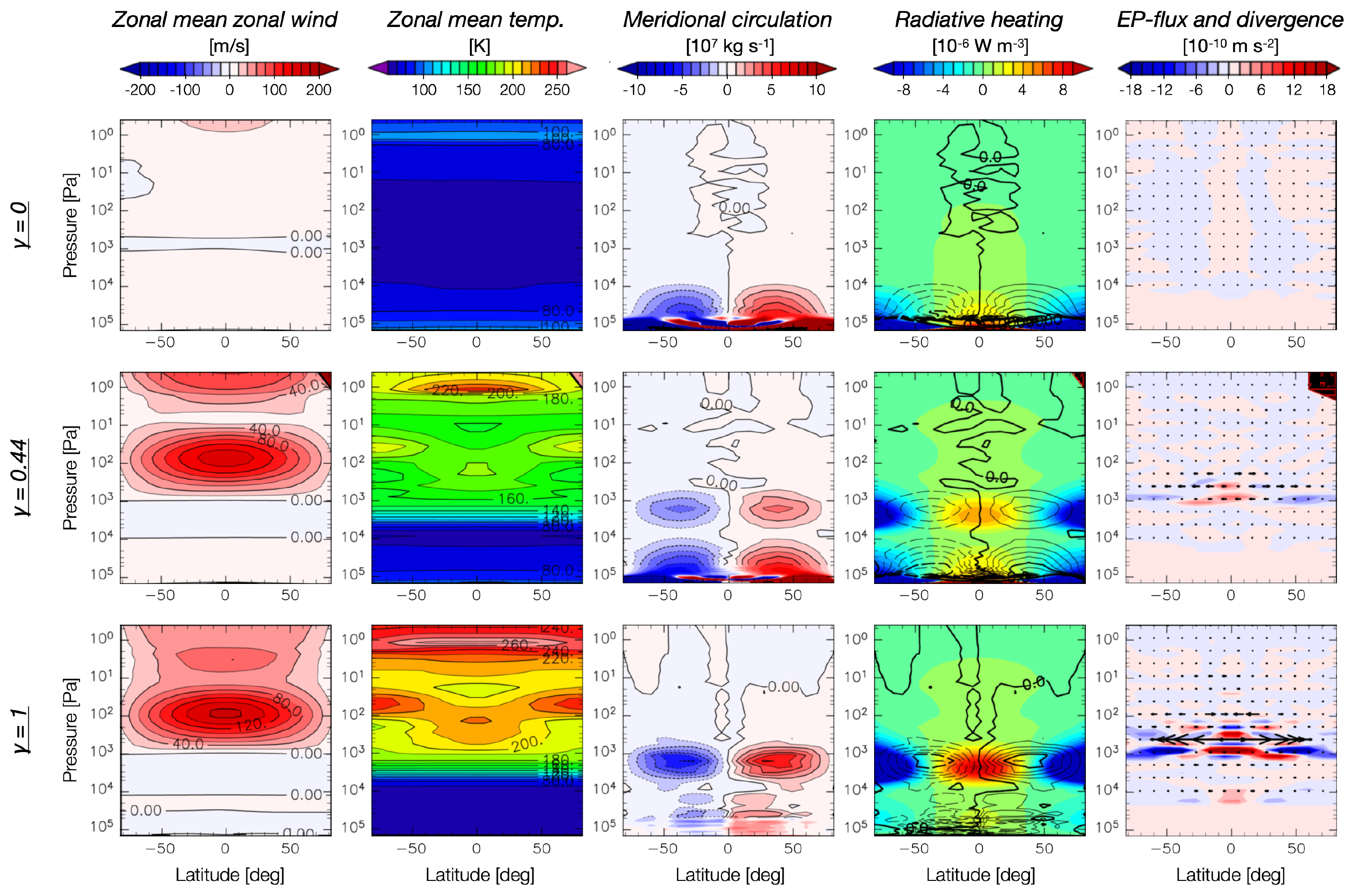}
    \caption{
      Zonal mean fields for
      several values of $\gamma$ with $n=1.4$ 
      after $10^5$ days 
      \revise{of dynamical integration from the radiation-only calculation.
      Each field is produced by averaging 6 hours interval data over 2160 Earth days.
      }
      From left to right, zonal wind, temperature, Transformed-Eulearian Mean mass stream function, radiative heating, and EP flux and its divergence are shown,  respectively. 
      Top, middle and bottom panels indicate
      the cases with $\gamma = 0,0.44$ and $1$,
      respectively.
      }
    \label{fig:summary_gamma}
\end{figure}

\begin{figure}[htbp]
    \centering
    \includegraphics[width=0.25\linewidth]{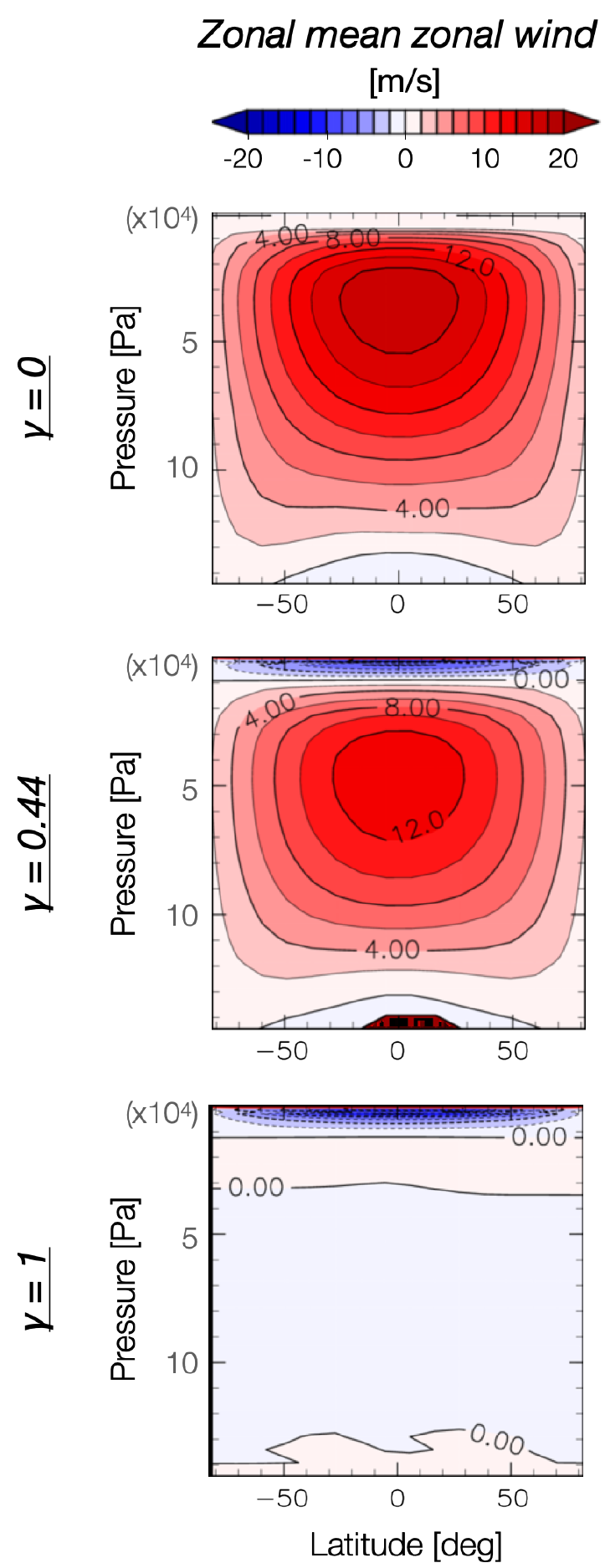}
    \caption{Zoom-in of the left column of \Figref{fig:summary_gamma} to show the zonal wind profile in the lower layers. 
    The top, middle, and bottom panels correspond to the case with $\gamma = 0,0.44$ and $1$, respectively.
    }
    \label{fig:summary_gamma_lower}
\end{figure}

\Figref{fig:summary_gamma} summarizes the meridional (latitude-pressure) structure of our experiments with varying $\gamma $. 
The middle row with $\gamma = 0.44$ corresponds to the most Titan-like temperature profile (see the second-left column), while $\gamma = 0$ (top row) and $\gamma = 1$ (bottom row) represents an atmosphere without a haze layer and that with a more significant haze layer, respectively. 

When the haze parameters are tuned for Titan ($\gamma = 0.44$; middle row), the global superrotating zonal flows appear both in the upper (stratosphere) and lower (troposphere) layers.
The eastward zonal wind above $10^3$ Pa pressure level is particularly strong with a maximum wind speed of about $140$~m/s at $10^2$~Pa. 
Zooming in the troposphere (\Figref{fig:summary_gamma_lower}) shows that the eastward zonal wind below $10^4$~Pa pressure level is much weaker with the maximum wind speed of $12$ m/s at $5\times 10^4$~Pa in the equatorial region. 
These amplitudes of the wind velocity in the upper and lower regions are comparable to the Huygens DWE data \citep{Bird2005}. Between these two superrotating regions, there is a layer where the wind is weakly westward, located around $3\times 10^4$~Pa, or $\sim $70~km in altitude. This altitude approximately corresponds to the location of the ``zonal wind collapse'' region. We discuss the formation of this layer in more detail in the following subsections. 

When $\gamma $ is varied, a clear trend in the stratospheric superrotation is seen. 
With $\gamma = 0$, i.e., when the solar flux is not absorbed by the haze layer, the stratospheric superrotation is substantially smaller than the $\gamma = 0.44$ case, while the weak superrotation in the lower layer remains. 
\Figref{fig:summary_gamma_lower} reveals that the zonal mean zonal wind in the lower layer is larger than the case of $\gamma = 0.44$, with the maximum wind speed of $16$ m/s. 
On the other hand, when $\gamma $ is increased to 1, i.e., when the solar flux is completely absorbed by the haze layer, the stratospheric superrotation becomes more prominent, with the maximum wind speed exceeding 160~m/s (\Figref{fig:summary_gamma}). 
The eastward wind in the lower layer disappears in this case (\Figref{fig:summary_gamma_lower}). 
These results indicate that the strong superrotation zonal flows in the upper layer are caused by the absorption of solar radiation by the haze layer. 

The second-left column of \Figref{fig:summary_gamma} shows the map of the zonal mean temperature in the meridional plane, which shows that the stratosphere temperature increases as $\gamma $ increases, consistent with the radiative equilibrium temperature profile  (\Figref{fig:initial_temperature}). 

The Transformed-Eulerian Mean (TEM) meridional circulation and the heating rate map are shown in the center and the second-right columns of \Figref{fig:summary_gamma}. 
More specifically, the center column represents the mass stream function averaged over the zonal residuals. 
In the case of $\gamma =0.44$ (middle row), the meridional circulations appear both in the lower and the upper layers, both of which rise around the equator and falls around the poles. 
Overlaying the map of the heating rate (the second-right column) reveals that the the upper circulations are produced by the latitudinal variation in the solar radiation absorbed by the haze layer, while the lower circulations are the latitudinal deviation of solar heating reaching at the planetary surface. 
Note that the temperature structure is almost uniform horizontally despite the differential heating along the latitude, because the meridional circulation compensates. 
Clearly, the heating rate and the strength of the resultant circulation is strongly affected by the value of $\gamma $. 
When the solar radiation is not absorbed by the haze layer ($\gamma = 0$), the meridional circulation appears only near the surface, while when the solar radiation is completely absorbed by the haze layer ($\gamma = 1$) the circulation develops predominantly in the upper layer. 
It is also evident that the strength of stratospheric superrotation is correlated with the strength of the meridional circulation. 

These results, together with the fact that the altitude of the base of the stratospheric superrotation coincides with the upper branch of the meridional circulation in the upper layer, are suggestive of the Gierasch mechanism. 
In Gierasch mechanism, the angular momentum that is transported from the lower layer by the meridional circulation is converged toward the equatorial region by the effective viscosity, generating eastward zonal flow. 
In order to identify the processes responsible for the effective viscosity, the detailed structure at this altitude is analyzed in the next subsection.

\subsubsection{Wave analysis}

In this section, we discuss the mechanisms that maintain stratospheric superrotation in the presence of the haze layer. 
Our analysis is based on the Transformed-Eulerian Mean (TEM) equation of the system that divides the acceleration of zonal flows by residual mean meridional flows and that by the disturbances (eddies). 
In TEM, the zonal component of the equation of motion is written as follows:
\begin{align}
  \dfrac{\partial \overline{u}}{\partial t}
  = \underbrace {-\dfrac{\overline{v}^*}{a \cos{\varphi}} \dfrac{\partial}{\partial \varphi}(\overline{u} \cos{\varphi}) - \overline{w}^* \dfrac{\partial \overline{u}}{\partial z^*} + f \overline{v}^*}_{\mbox{\textcircled{\scriptsize 1}}}
  + \overline{\mathcal{F}_\lambda}
  + \underbrace {\dfrac{1}{\rho_0 a \cos{\varphi}} {\bm \nabla \cdot \bm F}}_{\mbox{\textcircled{\scriptsize 2}}}
  \label{eq:uacc of TEM}
\end{align}
where $a$ is the planetary radius,  $\lambda$, $\varphi$ and $z$ are longitude, latitude and altitude, respectively, $\overline{(\cdot)}$ denotes zonal average. 
$\overline{u}$ is zonal mean longitudinal velocity, $\overline{v}^*$ and  $\overline{w}^*$ are the latitudinal and vertical components of residual mean meridional circulation, respectively. 
$f$ is the Coriolis parameter, and $\overline{\mathcal{F}_\lambda}$ is horizontal diffusion for numerical stability. 
$\rho_0$ is density.
\textcircled{\scriptsize 1} refers to the acceleration term of the zonal wind by the residual mean meridional circulation
$(0,\overline{v}^*,\overline{w}^*)$, 
which is described as, 
\begin{align}
    \overline{v^*} &=\overline{v} - \dfrac{1}{\rho_0}\dfrac{\partial}{\partial \overline{z}}
    \left(\rho_0\dfrac{\overline{v'\theta'}}{\frac{\partial \overline{\theta}}{\partial \overline{z}}}\right),
    \\
    \overline{w^*} &= \overline{w} + \dfrac{1}{a\cos\varphi}\dfrac{\partial}{\partial \varphi}
    \left(\cos\varphi\dfrac{\overline{v'\theta'}}{\frac{\partial \overline{\theta}}{\partial \overline{z}}}\right).
\end{align}
\textcircled{\scriptsize 2} refers to that by the eddies. 
Here, ${\bm F}=(0,F_\varphi,F_z)$ is called Eliassen-Palm flux (EP-flux), which is defined as follows.
\begin{align}
  F_\varphi
  & = \rho_0 a\cos{\varphi}
    \left( \dfrac{\partial \overline{u}}{\partial \overline{z}}
       \dfrac{\overline{v'\theta'}}{\frac{\overline{\partial  \theta}}{\partial z}} - \overline{u'v'} \right),
  \\
  F_{z}
  &= \rho_0 a \cos{\varphi} \left( \left[ f - \dfrac{\frac{\partial \overline{u}\cos{\varphi}}{\partial \varphi}}{a \cos{\varphi}} \right] \dfrac{\overline{v'\theta'}}{\frac{\overline{\partial \theta}}{\partial z}} - \overline{u'w'}\right),
\end{align}
where $(\cdot)'$ denotes deviation from zonal average, and $\theta$ is potential temperature. 
The EP-flux means the negative angular momentum transport due to non-zonal
disturbance fluid motions, and its divergence expresses acceleration
and deceleration of the mean flows by atmospheric waves.

The right column of \Figref{fig:summary_gamma} represents the EP-flux and its divergence. 
The wave activities are seen near $10^3$ Pa pressure level only when the radiative heating by the haze layer is active ($\gamma =0.44,\,1$). 
These EP-fluxes direct poleward, meaning positive angular momentum transport from the polar regions to the equatorial region. 
The EP-flux divergence fields also indicate that the equatorial regions are accelerated, suggesting that equatorial super-rotation states in the upper layer in these cases are maintained by the angular momentum transport of non-zonal wave motions.

\begin{figure}[htbp]
    \centering
    \includegraphics[width=\linewidth]{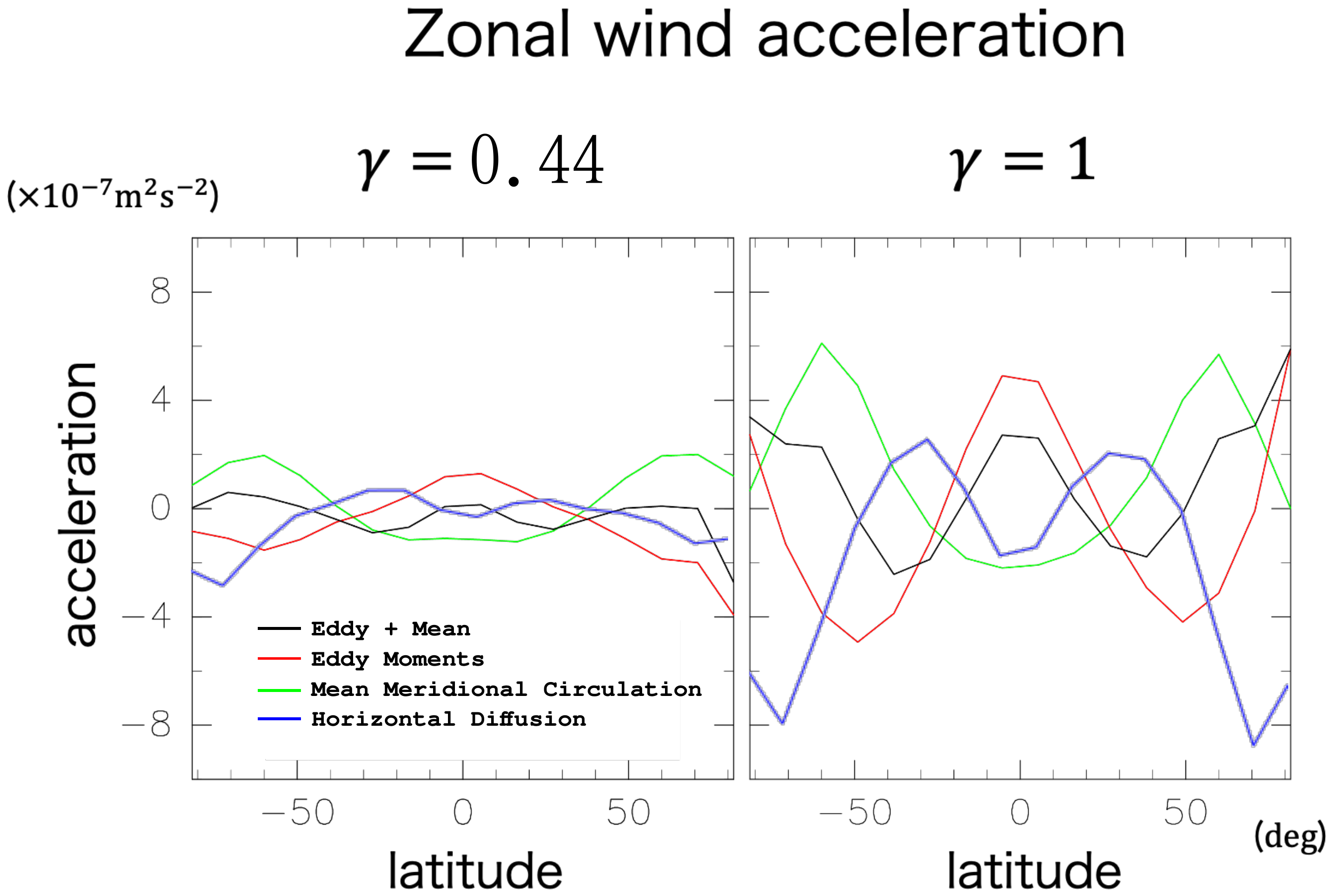}
    \caption{Latitudinal distributions of the terms in the zonal component of TEM momentum equation
      (\Equref{eq:uacc of TEM}) at $784$ Pa altitude.
      The red, green and blue lines represent acceleration by zonal mean residual meridional circulation
      (\textcircled{\scriptsize 1} in \Equref{eq:uacc of TEM}), 
      that by eddies (\textcircled{\scriptsize 2} in \Equref{eq:uacc of TEM}) , 
      \revise{and by horizontal diffusion},  respectively.
      The black lines represent total acceleration by eddies and mean meridional circulation (\textcircled{\scriptsize 1} $+$ \textcircled{\scriptsize 2}).
      Left and right panels are for $\gamma = 0.44$ and $1$,
      respectively.}
    \label{fig:TEMu}
\end{figure}

A closer look at the angular momentum transport at the $784$ Pa pressure level (where the wave activities are most prominent) in the cases of $\gamma = 0.44$ and $1$ is shown in \Figref{fig:TEMu}; The plots show the budgets of the angular momentum transport by the mean circulation and eddies. 
In both cases, the eddy term contributes to acceleration at the equator, while the mean field contributes to deceleration there. 
In contrast, at mid-latitudes ($\sim 50^{\circ }$), the mean flow contributes to acceleration, and the eddy terms contributes to deceleration. 
This means that the eddies transport the positive angular momentum from the latitude of $\sim 50^{\circ }$ to the equatorial regions, while the meridional circulation transport the angular momentum in the opposite direction. 
Note that the total acceleration by the eddies and mean flows is almost balanced with the deceleration by horizontal diffusion, resulting in a quasi-steady state.

\begin{figure}
  \centering
  \includegraphics[height=0.25\textwidth]{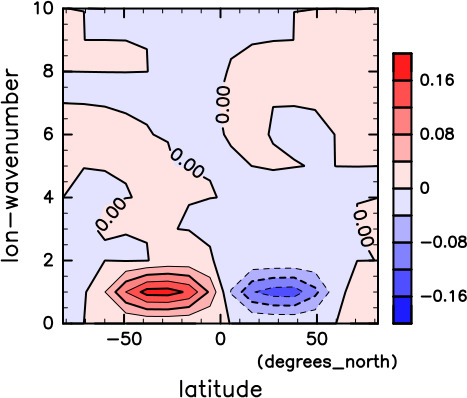}
  \includegraphics[height=0.25\textwidth]{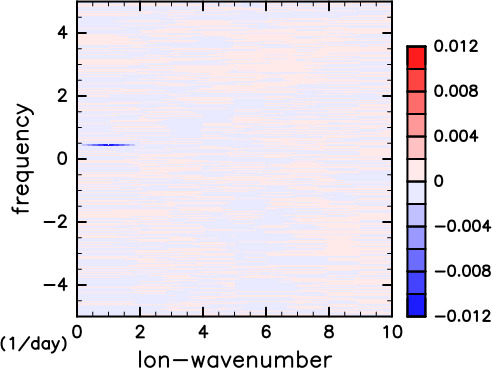}
  \includegraphics[height=0.25\textwidth]{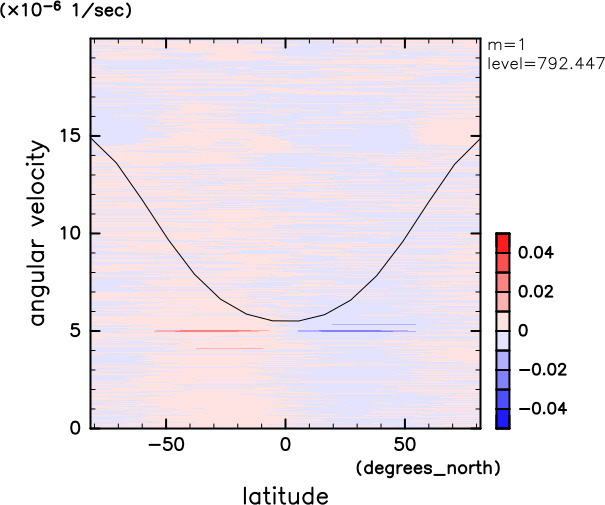}

  \caption{
    Spatio-temporal spectral analysis for the case with $\gamma=1$
    at $p=784$ Pa.
    The left panel indicate contribution of each longitudinal wavenumber component added up over the frequency to the co-spectrum of the longitudinal and latitudinal velocities. 
    The center panel shows the co-spectrum at $40^\circ$ north latitude. 
    The right panel is the co-spectrum and mean zonal angular velocity for $m=1$ components. 
  }
  \label{fig:UVcospectrum}
\end{figure}

In order to identify the wave that contributes to the equatorial acceleration most, we performed the co-spectrum analysis of the longitudinal and latitudinal velocity components in the case of $\gamma = 1$ at the altitude of $784$ Pa. 
The left panel of \Figref{fig:UVcospectrum} shows the  co-spectrum added up over the frequency at this altitude as a function of latitude and the longitudinal wavenumber $m$. 
The equatorward angular momentum transport of the eddies with wavenumber $m=1$ between latitude $0^{\circ }-50^{\circ }$ is prominent. 
The co-spectrum at the same altitude and at 40$^{\circ }$ latitude is shown in the middle panel of \Figref{fig:UVcospectrum}, in the plane of the longitudinal wavenumber $m$ and frequency $\omega$, revealing the predominant contribution of the component with $m=1$ and $\omega=0.432$ day$^{-1}$. 
The right panel compares the angular velocity of the $m=1$ component of co-spectrum and the mean zonal angular velocity as a function of latitude. 
The peak at the angular velocity ($\omega/m$) $5\times 10^{-6}$ sec$^{-1}$ corresponds to the component with $\omega=0.432$  day$^{-1}$. 
The fact that the angular velocity is smaller than the mean zonal angular velocity suggests the Rossby-wave nature.

%

\begin{figure}
  \centering
  \includegraphics[width=0.8\linewidth]{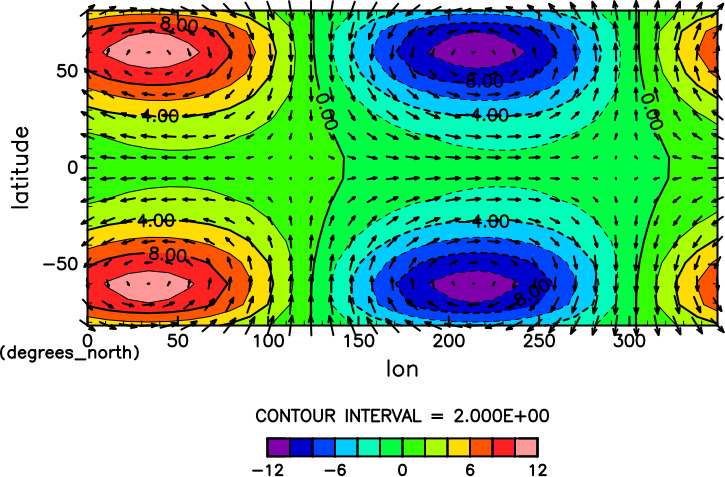}
  \caption{
    Horizontal winds and geopotential height
    of the wave with zonal wavenumber 1
    and frequency 0.432 $({\rm day^{-1}})$
    at an altitude of $784$ Pa for the case of $\gamma=1$.
    The horizontal axis is longitude and the vertical axis is latitude.
    The arrows indicate horizontal wind field,
    and tone indicates geopotential height,
    where red and blue colors correspond to high and low pressures,
    respectively.
  }
  \label{fig:UVH}
\end{figure}

The global structure of this wave
with $m=1$ and $\omega=0.432$ day$^{-1}$ is shown in \Figref{fig:UVH}. 
It can be seen that the horizontal wind is blowing
along the isobaric contours over the entire globe,
especially from mid to high latitudes. 
In addition, the isobaric contours have a tilting structure
toward the eastward and equatorward.
This structure indicates transport of eastward angular momentum toward the equator, confirming that this wave is responsible for the equatorial acceleration. 
The dominant vortices in high-latitudes and the relative structure between the velocity field and the geopotential height again indicate that this wave is a Rossby wave. 

\begin{figure}
  \centering
  \includegraphics[height=0.3\textwidth]{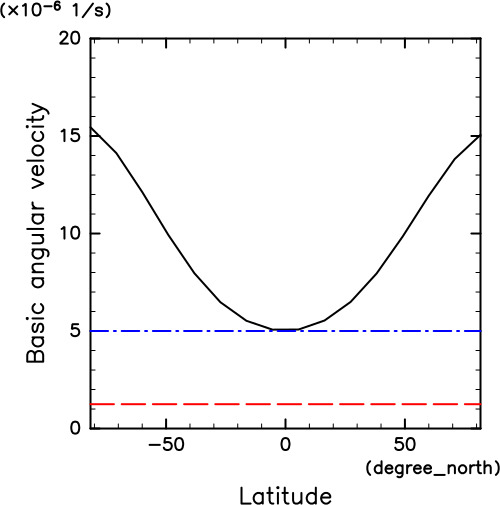}
  \includegraphics[height=0.3\textwidth]{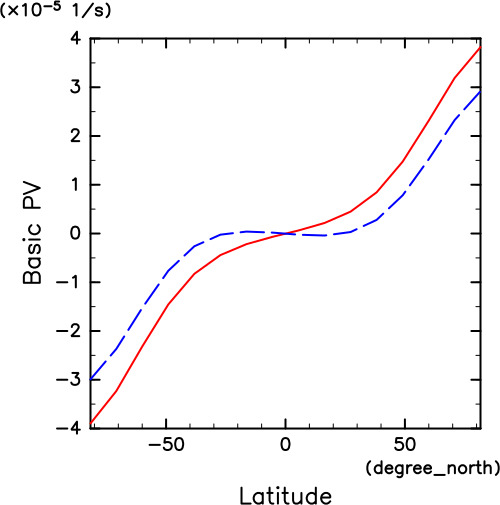}
  \vspace{0.3cm}
  
  \includegraphics[height=0.3\textwidth]{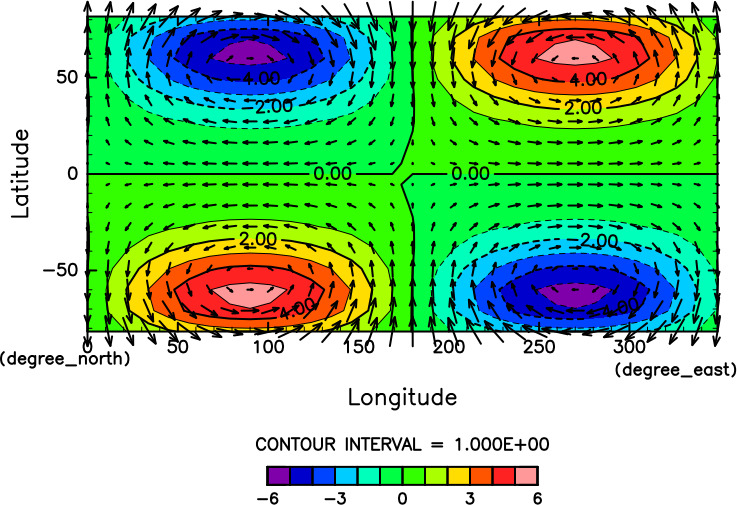}
  \includegraphics[height=0.3\textwidth]{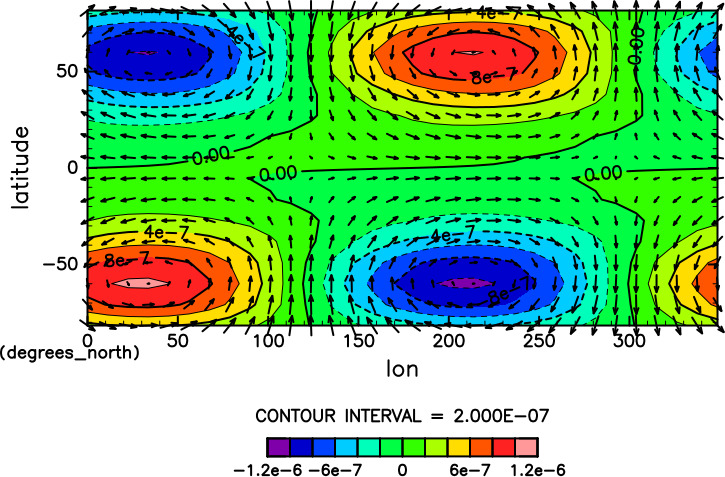}
  \caption{
    Linear stability analysis of a 1-layer barotropic model with horizontal diffusion. 
    The upper left panel shows
    angular velocity of the zonal mean zonal wind of 
    the basic state for the stablity analysis (black solid line), frequency of the unstable mode (red broken line)
    and frequency of the wave derived from GCM simulation (blue dotted line). 
    In the upper right panel, 
    the blue broken line and red solid line indicate zonal mean vorticity and potential vorticity (absolute vorticity) of the basic state. 
    The lower left panel is vorticity and horizontal wind fields of the unstable mode with the growth rate and frequency of $2.5\times 10^{-9}$ and $1.25\times 10^{-6}$, respectively.
    The vorticity and horizontal wind fields obtained by the wave analysis are shown in the lower right panel for comparison. }
  \label{fig:rotbarolinear}
\end{figure}

The excitation mechanism of the dominant wave may be barotropic instability
at the 800Pa level.
Kelvin or gravity waves would not take part in this instability. 
We performed linear stability analysis of the basic state derived from the same GCM result 
(the upper panels of Fig.\ref{fig:rotbarolinear}) using the two dimensional barotropic vorticity equation on a rotating sphere, 
and found an unstable mode with the longitudinal wave number 1 (the lower left panel of Fig.\ref{fig:rotbarolinear}). 
Its structure is similar to that obtained by the wave analysis of GCM result
(the lower right panel of Fig.\ref{fig:rotbarolinear}), 
although the frequency of the unstable mode is rather lower than that by the wave analysis\revise{, possibly because the basic state is not an exact solution of the axisymmetric governing equations but also includes the effects of wave components}.
The same horizontal diffusion term for vorticity as that used in the GCM is introduced for the analysis. 
No unstable mode is observed when the horizontal diffusion is switched off. 
Note that there is no inflexion point where the latitudinal gradient of basic potential vorticity vanishes nor a critical latitude where the phase velocity coincides with the mean angular velocity of the background, which is in contrast with the necessary conditions for inviscid  barotropic instability. 
However, viscous flows may be able to be  unstable without satisfying the inviscid stability criterion.
For example, viscous Poiseuille flow becomes unstable for a certain range of the Reynolds number. A discussion by the
integrals of Orr-Sommerfeld equation indicates relaxation of the existence of a critical point for unstable modes of viscous parallel flows \citep{Drazin-Reid2004}.

\subsubsection{The Gierasch mechanism detached from the surface}

While we confirmed the structure reminiscent of the Gierasch mechanism, namely the meridional circulation and the equatorward angular momentum transport in the upper part of the circulation cell, there is a difference from the original Gierasch mechanism. 
In the original Gierasch mechanisms, the meridional circulation is connected to the surface and the source of angular momentum is ultimately the solid surface. 
However, in our experiments with the haze layer ($\gamma = 0.44$ and $1$), the meridional circulation is largely detached from the surface. 
In such a situation, where is the source of the angular momentum of the superrotation? 

\begin{figure}
  \centering
  \includegraphics[height=0.4\linewidth]{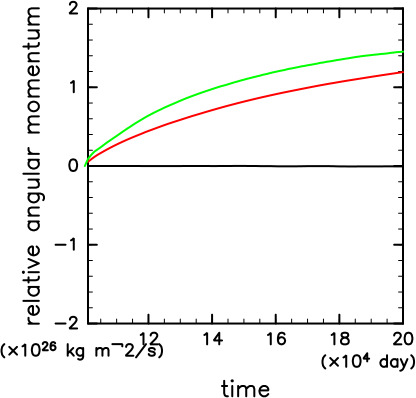}
  \includegraphics[height=0.4\linewidth]{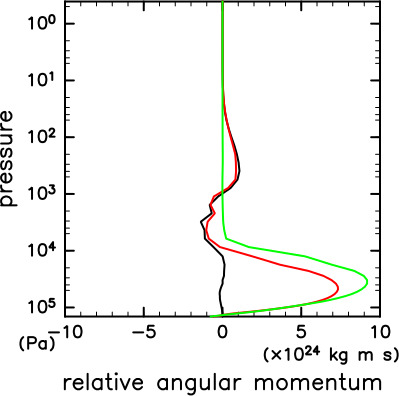}
  \caption{
    Time development of total relative angular momentum (left panel)
    and vertical distributions of horizontally integrated
    relative angular momentum (right panel). 
    Black, red and green lines represent
    the $\gamma=1$, $0.44$ and $0$ cases, respectively. 
  }
  \label{fig:total-angular-momentum}
\end{figure}

The left panel of \Figref{fig:total-angular-momentum} shows
time development of the globally integrated relative angular momentum for varying $\gamma$. 
When $\gamma = 0$ or $0.44$, the relative angular momentum is positive and increasing during this timeframe, indicating that the atmosphere is extracting the positive angular momentum from the solid surface. 
In contrast, for the case with $\gamma=1$, the integrated relative angular momentum nearly equals to zero, meaning that angular momentum is redistributed {\it within} the atmosphere, without the exchange of angular momentum with the solid planet. 

The right panel of \Figref{fig:total-angular-momentum} shows vertical distributions of horizontally integrated relative angular momentum at $10^5$ Earth days. 
For the case with $\gamma=1$, positive and negative deviations of relative angular momentum above and below the level at $10^3$ Pa, respectively, and they are nearly balanced each other. 
This indicates that the strong westerly winds above $10^3$ Pa level is produced by extracting positive relative angular momentum from just below the layer. 
A similar accumulation of the positive and negative relative angular momentum in the upper layer is also found when $\gamma=0.44$, suggesting that at least some of the acceleration above $10^3$ Pa level is transported from the slightly lower region in the atmosphere ($p \sim 3\times 10^3$ Pa). 
%

\subsection{Dependence on the altitude of haze layer}

\begin{figure}[htbp]
    \centering
     \includegraphics[width=\linewidth]{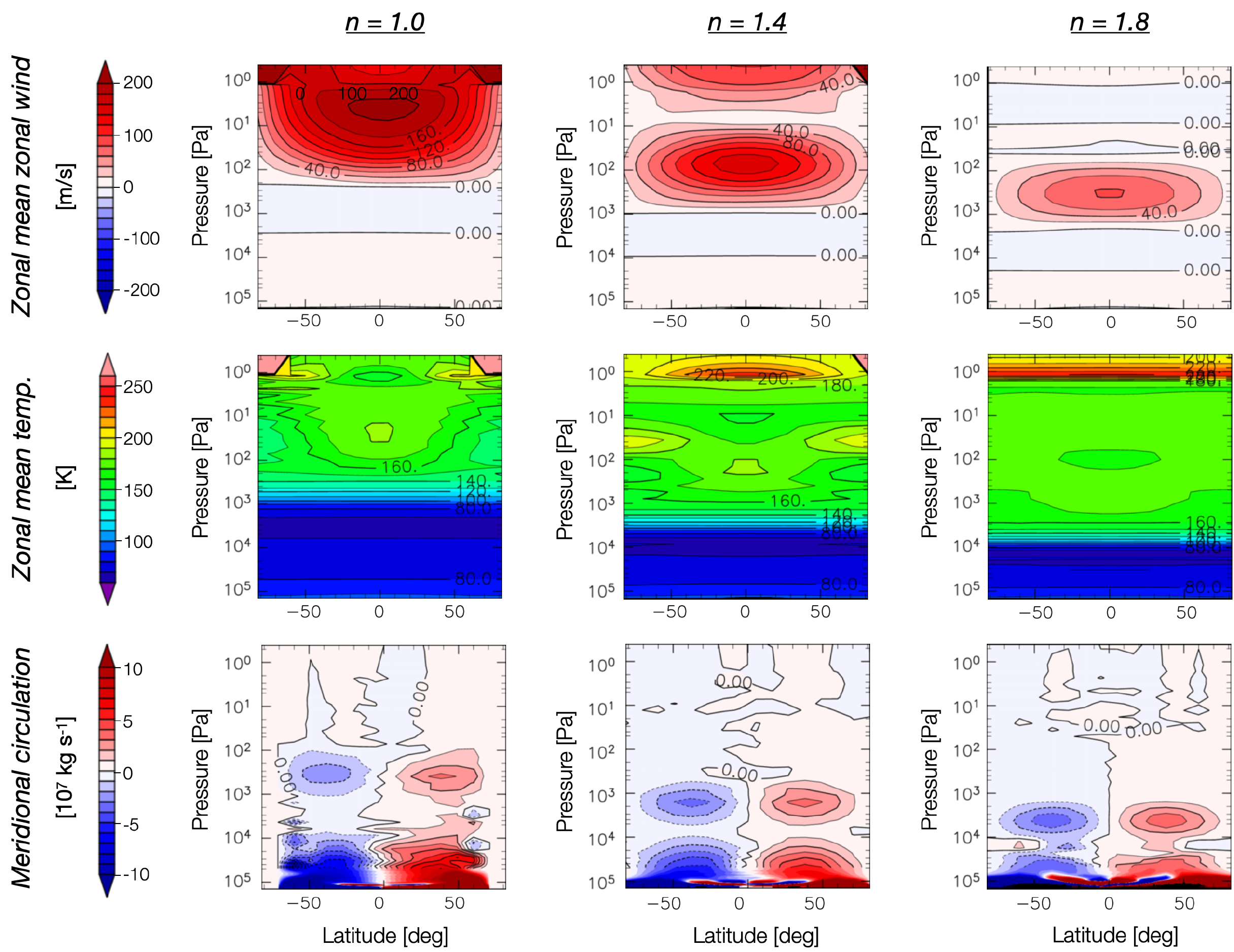}
    \caption{
      Zonal mean fields 
      for several values of $n$ with $\gamma=0.44$
      \revise{after $10^5$ days of dynamical integration from the radiation-only calculation.
      Each field is produced by averaging 6 hours interval data over 2160 Earth days.}
      Top, middle and bottom panels show zonal wind, temperature, and TEM mass stream function, \revise{respectively}. 
      Left, center and right panels indicate
      the cases with $n= 1, 1.4$ and $1.8$, respectively.
    }
    \label{fig:summary_n}
\end{figure}

Next, we show the dependence of the zonal mean meridional fields
on the altitude where the solar radiation is effectively absorbed
by the haze layer ($n$). 
\Figref{fig:summary_n} shows the $n$-dependence of zonal mean zonal wind fields, zonal mean temperature, and meridional circulation. 
Note that the rows and columns are different from \Figref{fig:summary_gamma}; the columns corresponds to different values of $n$. 
Clearly, the change in $n$ changes the altitude of the strong superrotation, that of the largest temperature gradient, and that of the meridional circulation in the upper layer, in a consistent manner. 
In all cases, the relative altitude among them are similar, suggesting the consistent underlying process of generating superrotation. 
Note that the zonal wind velocity and horizontal temperature differences decrease as the altitude of the superrotation is lowered, 
while the amplitude of residual mean circulation does not vary very much. 
This is probably due to the higher density at lower altitude since the acceleration strength of circulation per unit volume by the effect of the haze layer seems to be similar in all cases.

\subsection{Result of longer time integration}

\begin{figure}[htbp]
    \centering
    \includegraphics[width=0.8\linewidth]{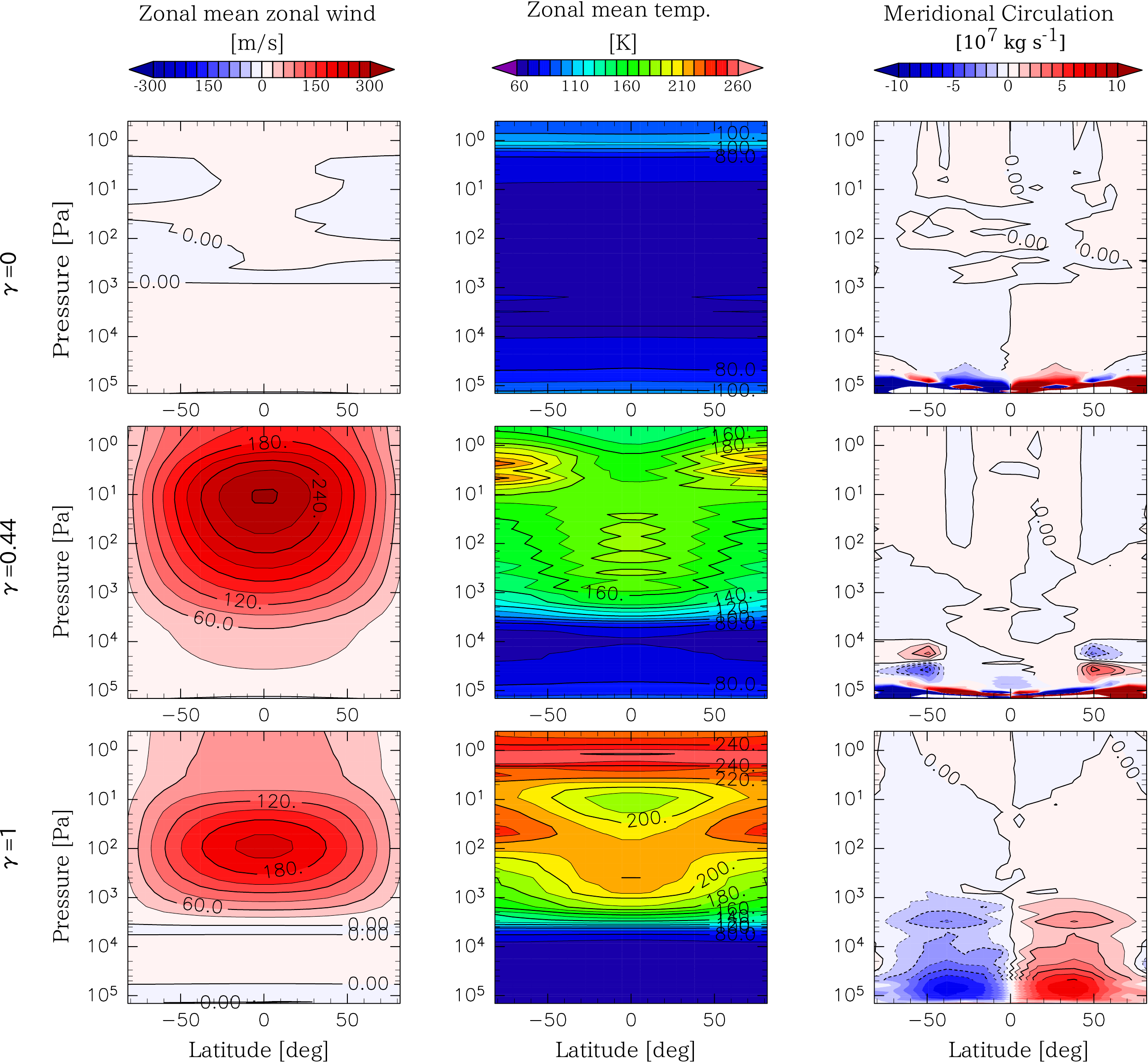}
    \caption{
      Zonal mean fields after $10^6$ days 
      \revise{of dynamical integration from the radiation-only calculation.
      Each field is produced by averaging 6 hours interval data over 2160 Earth days.}
      From left to right, zonal wind, temperature, mass stream function are shown,  respectively. 
      Top, middle and bottom panels indicate
      the cases with $\gamma = 0,0.44$ and $1$,
      respectively.
      }
    \label{fig:summary_gamma_t1e6}
\end{figure}

The results with the radiation parameters with $\gamma=0.44$, $1$, and $n=1.4$ discussed above are not in a full dynamical equilibrium
after $10^5$ days from radiative-only calculation. 
Positive angular momentum still accumulates (left panel of 
\Figref{fig:total-angular-momentum}),
and eastward wind develops gradually in the troposphere. 
Thus, we performed further time integrations until $10^6$ days, 
the results of which are summrized in Fig. \ref{fig:summary_gamma_t1e6}.
Stratospheric superrotation evolves both in the case with $\gamma=0.44$ and $1$, 
the maximum amplitudes of which reach $300$ and $200$ m/s, respectively. 
When $\gamma=0.44$, 
the westward wind region just below the stratospheric superrotation region disappears
after longer time integration
leaving a minor kink in the vertical gradient of the zonal wind, 
while it remains for $\gamma=1$ although the vertical thickness and amplitude are diminished. 
A similar time evolution can be seen in the long term simulation
of Venusian atmosphere performed by \citet{Sugimoto+2019}.
The meridional circulations at the haze layer are also weakend
both in the cases with $\gamma=0.44$ and $1$. 
The circulations near the surface for $\gamma=0.44$ 
show double cell structure in each hemisphere, 
while they are single cell but are strengthened 
for $\gamma=1$. 
Therefore, the features observed in the simulated results at $10^5$ days were transient. 
However, it is interesting that the simulation result at $10^5$ days well capture the observed zonal wind profile compared to that at $10^6$ days.
We will discuss this point in the summary and discussion section below. 

\subsection{Dependence on the horizontal resolution}

\begin{figure}[htbp]
    \centering
    \includegraphics[width=0.8\linewidth]{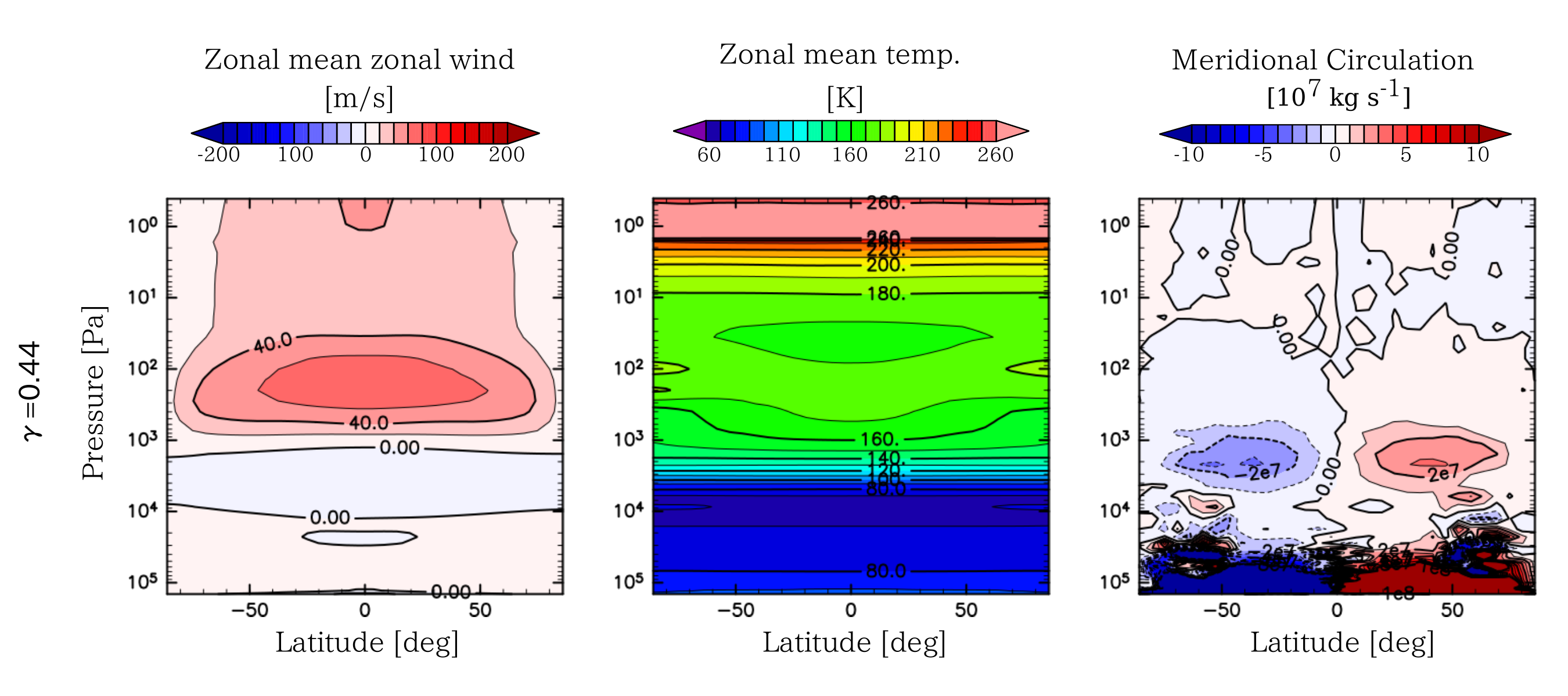}
    \caption{
      Zonal mean fields after $10^5$ days
      \revise{of dynamical integration from the radiation-only calculation} 
      for $\gamma = 0.44$ and $n=1.4$ 
      with the horizontal resolution of T21. 
      Each field is produced by averaging 6 hours interval data over 2160 Earth days.
      From left to right, zonal wind, temperature, TEM mass stream function are shown,  respectively. 
      }
    \label{fig:summary_G057_T21L55}
\end{figure}

Fig.~\ref{fig:summary_G057_T21L55} shows the zonal mean fields obtained by the simulation with higher horizontal resolution of T21
after $10^5$ days from the radiation-only calculation for $\gamma = 0.44$ and $n=1.4$. 
The characteristics of the meridional distributions of zonal mean zonal wind, zonal mean temperature and meridional circulations are qualitatively similar to those obtained by the T10 simulations (Fig.\ref{fig:summary_gamma}). The stratospheric superrotation wind emerges at the altitude of the haze layer. 
However, its maximum wind speed is only 80~m/s, which is weaker than that of the T10 calculation. 
Just below the superrotating wind region, there exists weak westward wind between $10^3$ and $10^4$ Pa levels. 
The meridional circulations appear both in the lower and the upper layers, 
which rise around the equator and fall around the poles.
Their location is similar to the T10 result but the amplitude and vertical thickness of the lower tropospheric circulations
are intensified. 
Similar distribution of zonal mean temperature is obtained although the latitudinal temperature contrast is weakened. 
These results suggest that the simulations with T10 resolution capture the robust features of the Titan's atmospheric circulation. 

\section{Summary and Discussion} \label{sec:conclusion}

In this study, we investigated the effects of atmospheric heating of the haze layer on the general circulation of the Titan's atmosphere in order to understand the mechanism responsible for the generation and maintenance of observed stratospheric superrotation.
A General Circulation Model (GCM), DCPAM, which has been developed by GFD Dennou Club based on the primitive equation system for planetary atmospheres,  was used for numerical experiments.
We employed the semi-gray radiation model of \citet[]{McKay1999}, where the absorption of the sunlight by the haze layer is parameterized as a function of pressure with the absorption rate $\gamma$ and the power-law index of optical depth $n$, which practically changes the altitude of the absorption 
in addition to the greenhouse effects by the atmospheric gases.
From an initial condition with a radiative-equilibrium temperature profile and no motion, 
the model was run for $10^5$ Earth days, which is approximately same as the radiative timescale near the surface for each experiment.

The numerical experiments with varying $\gamma$ and $n$ indicate that stratospheric heating by the haze layer is necessary to generate and maintain Titan's stratospheric superrotation, and that the vertical profile of the zonal wind is strongly affected by that of haze layer. 
Mean meridional circulations are driven at the altitude where the absorption of the haze layer is active, and equatorward angular momentum transport by waves along the upper branch of the meridional circulation brings back superrotating mean zonal flows to lower latitudes. 

A linear stability analysis with the two dimensional barotropic vorticity equation on a rotating sphere
suggests that the excitation mechanism of the dominant wave that transports the angular momentum to the equatorial region in the upper layers of the meridional circulation \revise{in the quasi-stationary state}
is barotropic instability affected by the horizontal diffusion at the 800 Pa level. 
It seems different from ageostrophic instability proposed by \cite{WangMitchell2014} for example, 
where the Kelvin and/or gravity waves play an important role. 
\revise{However, equatorial acceleration by ageostrophic instability may operate in the initial spin-up stage where faster mean zonal flows appear in high latitudes similar to those shown in \cite{MitchellVallis2010} for example. }
\revise{Note that the horizontal angular velocity diffusion plays important role in barotropic instability responsible for maintenance of superroation zonal wind in our simulations, 
which should be further examined whether the hyper diffusion type parameterization for the effects of sub-grid scale eddies is adequate for application to real Titan atmosphere. }

We performed simulations with higher horizontal resolution of T21, 
and confirmed that the characteristics of the meridional profiles of zonal mean zonal wind and other variables are qualitatively similar 
to those obtained by the T10 simulations.
\revise{
The successful results with T10 resolution may due to the fact that the dominant wave contributing the production of equatorial superrotaion has a global structure with longitudinal wavenumber one.  
However, there are significant quantitative differences between T10 and T21 simulations, especially for the amplitude of superrotating zonal winds. 
Similar dependence on the horizontal resolution is reported by the Venus GCM intercomparison work presented by \cite{Lebonnois2013}. 
Although our simulations with T10 resolution can capture the robust features of the Titan's atmospheric circulation, 
higher horizontal resolution would be necessary for quantitative discussion on the Titan's atmospheric dynamics. }

\begin{figure}[htbp]
    \centering
    \includegraphics[width=\linewidth]{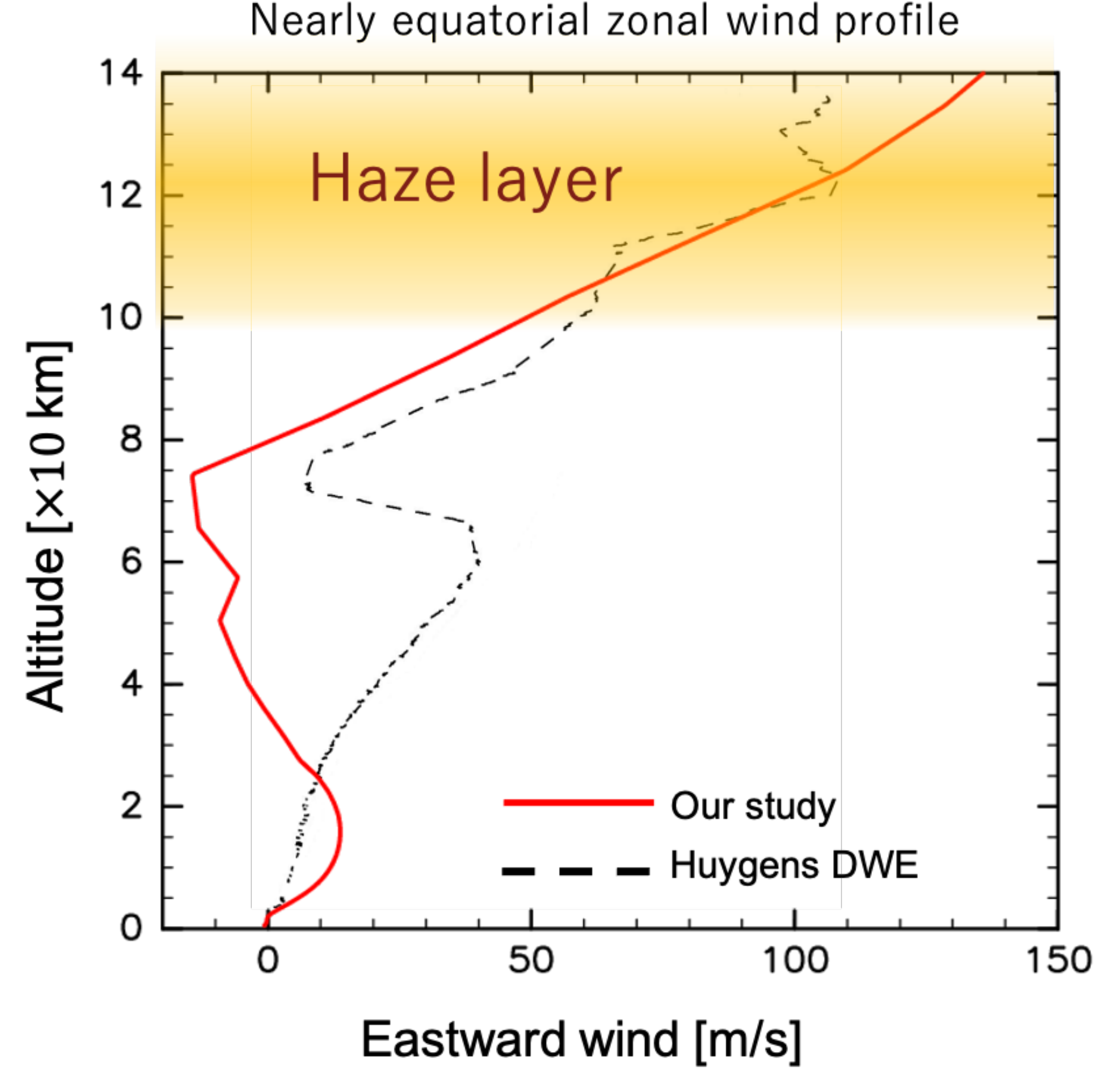}
    \caption{Comparison of zonal winds between observations
      and numerical experiments.
      The dotted line is longitudinal wind distribution at about $10\degree$S
      obtained by Huygens DWE,
      and the red solid lines is the result with $\gamma=0.44$ and $n=1.4$,
      which is the most preferable for the Titan's atmosphere
      in our numerical experiments 
      \revise{after $10^5$ days of dynamical integration from the radiation-only calculation.}
      }
    \label{fig:comparison_sim_obs}
\end{figure}

The numerical result with the radiation parameters 
$\gamma=0.44$ and $n=1.4$, which reproduce the vertical temperature
structure of Titan in a radiative equilibrium,
is not in a full dynamical equilibrium. 
%
%
Nevertheless, it is interesting to compare our results
with the zonal flow profile of Titan observed by DWE \citep{Bird2005}. 
In addition to the good agreement in the vertical shear winds above an altitude of 80 km, the observed ``zonal wind collapse'' at altitudes between 60 km and 100 km seems to correspond to the region of the negative angular momentum found in our simulation (Fig. \ref{fig:comparison_sim_obs}). 
This suggests that the ``wind collapse'' region in Titan's atmosphere may be a structure that is related to the fact that the seasonal cycle of the Titan (with the period of 30 Earth years) is shorter than the radiative relaxation time, which prevents the atmosphere from being fully equilibrated. 
Future simulations with seasonal variations would be needed to clarify this point. 
A comparison between the observed profile and our numerical results also shows other minor differences, including the sign of the zonal wind in the ``wind collapse'' region and the wind amplitude in the troposphere. 
The cause is not clear, but they may suggest that the processes that are not taken into account by our model, including the seasonal variation and the phase change of methane, plays a major role in determining the wind structure in the lower atmosphere. It is therefore to be studied whether the mechanism discussed in this paper really applies to the actual profile of Titan, 
by developing the model to a more realistic one in a step-by-step manner. 

It is worth noting that more realistic Titan GCMs \citep{Newman2011, Lebonnois2012, Lora2015} do not show the negative zonal wind with the stacked overturning circulation as seen in our results, 
\revise{although they do show a weaker minimum in zonal wind at around the right altitude.}
This difference may be explained by the difference in the integration time. 
Indeed, in our experiment with Titan-like parameters it appears only as a transient structure that is not seen in the equilibrium state after $10^6$~days. 
The zonal wind profile at the final state is closer to the profiles of those of the previous realistic simulations. 
There are also other differences in assumptions between our simplified model and the more realistic previous models. 
In addition to the inclusion of the seasonal variation and the phase change of methane, radiative transfer scheme and other numerical treatments differ. 
Adding individual elementary processes will be the scope of future work and will allow us to further study the causal relationship between each property of Titan and its atmospheric structure. 
%
%

\begin{figure}[htbp]
    \centering
    \includegraphics[width=\linewidth]{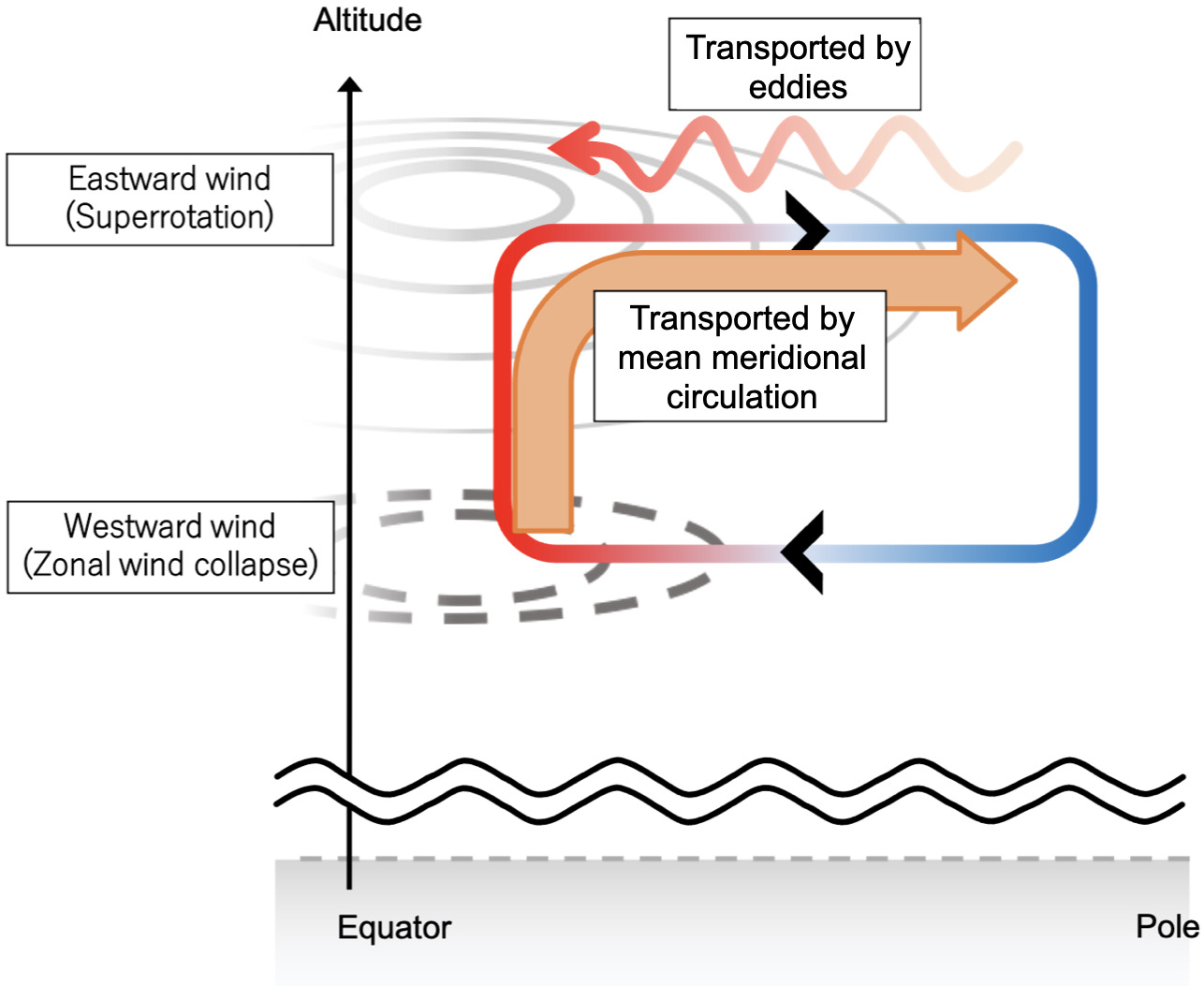}
    \caption{Schematic diagram of the mechanism suggested by our study to maintain stratospheric superrotation on Titan. There is a difference with \Figref{fig:Gierasch} regarding the presence of the surface.
      }
    \label{fig:newGierasch}
\end{figure}

Our results can also be compared with a series of GCM simulations
of the Venusian atmosphere performed by several authors
\citep[e.g.,][]{YamamotoTakahashi2003a, YamamotoTakahashi2003b, YamamotoTakahashi2004, YamamotoTakahashi2006, YamamotoTakahashi2009, KidoWakata2008, KidoWakata2009, Lebonnois+2010, Lee+2005, Lee+2007},  
which also obtained superrotation solutions.
Their standard experimental setup consists of
Newtonian cooling with equilibrium temperature
profiles, and includes localized heating around 60~km altitude
assuming absorption of incoming solar radiation by the cloud
layer.
After the long term integrations to spin up the atmospheres, 
the systems reach global superrotation states spreading
from the surface to the altitude of cloud layer. 
It seems that superrotation is a common feature
for the atmosphere of slowly rotating planets 
with localized heating in the upper atmosphere. 
However, their results usually show a monotonic increase
in zonal wind velocity up to the altitude of maximum wind velocity,
without deceleration region. 
While one of the possible causes for this difference 
would be the integration time, 
there is an interesting difference in relative heating and vertical static stability profile between Titan and Venus.
Typical configuration for the Venusian atmosphere used in their 
experiments brings relatively weak static stability
below the cloud layer.
Since the weak static stability promotes ageostrophic and/or
baroclinic instabilities 
\citep{WangMitchell2014, Sugimoto+2014a, Sugimoto+2014b, Kashimura+2019}
which may be preferable for equatorward angular momentum transport
by wave activities, the superrotation may emerge broadly
below the cloud layer.
In contrast, our results show that the weak stability layer spreads above the haze layer, 
where dominant superrotation emerges, and that
a strong stability layer exists just below the haze layer,
which seems to disconnect angular momentum transfer from the surface layer, 
resulting in the deceleration region (\Figref{fig:newGierasch}). 
This is a difference from the traditional Giearash mechanism, which pumps up the atmospheric angular momentum from the solid surface by the assumed infinite horizontal viscosity in the whole atmospheric layer. 
Further comparison and understanding of GCM simulation results of the atmospheres of Titan and Venus will give us more insights 
into generation and maintenance mechanism of superrotaion of Earth-like planets in general. 

\vspace{1em}






We acknowledge the constructive comments from the anonymous reviewers, which helped us improve this manuscript. 
DCPAM5 is used as the numerical model for this study. The figures and analysis were produced by GFD Dennou Ruby project\footnote{http://ruby.gfd-dennou.org}.
Numerical computations were in part carried out on Cray XC50 at Center for Computational Astrophysics, National Astronomical Observatory of Japan.
YF is supported by \revise{Grant}-in-Aid from MEXT of Japan, No. 18K13601. ST is supported by \revise{Grant}-in-Aid from MEXT of Japan, No. 21H01155.


\appendix

\section{Equations for DCPAM5}
\label{ap:dcpam5}

In this section, we summarize the equation system used in our GCM, DCPAM5. 

To begin with, we write down the primitive equations with respect to the following coordinates:
\begin{align*}
\varphi&: \mbox{latitude}, \\
\lambda&: \mbox{longitude}, \\
\sigma &\equiv p/p_{\rm surf}, \\
t&: \rm \mbox{time},
\end{align*}
where $p$ is the pressure, and $p_{\rm surf}$ is the surface pressure. 
The total derivative is expressed as
\begin{equation}
\dfrac{d}{dt} \equiv \dfrac{\partial}{\partial t} + \dfrac{u}{a\cos{\varphi}}\dfrac{\partial}{\partial \lambda} + \dfrac{v}{a} \dfrac{\partial}{\partial \varphi} +\dot{\sigma}\dfrac{\partial}{\partial \sigma}.
\end{equation}

The main variables to solve are
\begin{align*}
p_{\rm surf}&: \mbox{surface pressure,} \\
(u,v,\dot \sigma)&: \mbox{velocity vector for } (\lambda, \varphi, \sigma ) \mbox{ coordinates,} \\
T&: \mbox{temperature.}
\end{align*}
For convenience, we also use the following auxiliary notations.
\begin{align*}
\bm{v}_H &= (u, v) \mbox{   (horizontal wind velocity)} \\
\pi &\equiv \ln{p_{\rm surf}}, \\
\Phi &\equiv gz: \mbox{geopotential},
\end{align*}
where $g$ is the gravity and $z$ is the altitude.

Using these expressions, each equation of the primitive equations is described below.

\paragraph{The continuity equation}

\begin{align}
\dfrac{d \pi}{d t} + \nabla \cdot \bm{v}_H + \dfrac{\partial \dot{\sigma}}{\partial \sigma} = 0.
\end{align}
Note that the vertical velocity in $\sigma$ coordinates,  $\dot{\sigma}$, can be written as
\begin{align}
\dot{\sigma} \equiv \dfrac{g\sigma}{R^dT}\left\{ \left( \dfrac{\partial z}{\partial t}\right)_\sigma + \dfrac{u}{a \cos{\varphi}} \left( \dfrac{\partial z}{\partial \lambda }\right)_\sigma + \dfrac{v}{a} \left( \dfrac{\partial z}{\partial \varphi}\right)_\sigma -w \right\},
\end{align}
where $z$ is the altitude, $a$ is the planet radius and $w$ is the radial velocity with respect to $z$.
The $R^d$ is defined by $R^d \equiv R/\overline{M}$ with $R$ being the gas constant  and $\overline{M} $
being the mean molecular weight of air.

\paragraph{The hydrostatic equation}

\begin{align}
\dfrac{\partial \Phi}{\partial \sigma} = - \dfrac{R^dT}{\sigma},
\end{align}
which adopts the equation of state for ideal gas.

\paragraph{The equations of motion}

\begin{align}
\dfrac{d u}{d t} - fv - \dfrac{uv}{a}\tan{\varphi} &= - \dfrac{1}{a\cos{\varphi}} \dfrac{\partial \Phi}{\partial \lambda} - \dfrac{R^dT}{a\cos{\varphi}} \dfrac{\partial \pi}{\partial \lambda} +\mathcal{F}_\lambda,  \label{eq:motion of u} \\
\dfrac{dv}{d t} + fu + \dfrac{u^2}{a}\tan{\varphi} &= -\dfrac{1}{a}\dfrac{\partial \Phi}{\partial \varphi} - \dfrac{R^dT}{a} \dfrac{\partial \pi}{\partial \varphi} + \mathcal{F}_\varphi, \label{eq:motion of v}
\end{align}
where $f\equiv 2\Omega \sin \varphi$ is the Coriolis parameter with the angular velocity $\Omega $, while $\mathcal{F}_\lambda$ and $\mathcal{F}_\varphi$ are the external forces in the direction of $\lambda $ and $\varphi $, respectively.

\paragraph{The thermodynamic equation}

\begin{align}
\dfrac{d T}{d t} = \kappa T\left\{ \dfrac{\partial \pi}{\partial t} + \bm{v}_H \cdot \nabla_\sigma \pi + \dfrac{\dot{\sigma}}{\sigma} \right\} +\dfrac{Q}{C_p},
\end{align}
where $\kappa \equiv R/C_p$ with $C_p$ being the specific heat of air at constant pressure and
$Q$ is the radiative heating given by \Equref{eq:Heating rate}.

The primitive equations used in DCPAM5 are a variant of the set of equations introduced above, with the following transformation of variables. 
Let us first transform the spatial coordinate by replacing $\varphi $ by $\mu $:
\begin{equation}
    \mu \equiv \sin{\varphi}.
\end{equation}
Then, the horizontal velocities are also transformed as follows for easier computations on a sphere:
\begin{align}
U &\equiv u \cos{\varphi} = u \sqrt{1-\mu ^2}, \\
V &\equiv v \cos{\varphi} = v \sqrt{1-\mu ^2}.
\end{align}
Furthermore, we introduce Vorticity, $\zeta$, and Divergence, $D$:
\begin{align}
\zeta &\equiv \dfrac{1}{a\cos{\varphi}}\dfrac{\partial v}{\partial \lambda} - \dfrac{1}{a\cos{\varphi}}\dfrac{\partial}{\partial \varphi}(u\cos{\varphi}) \notag \\
&=\dfrac{1}{a(1-\mu^2)}\dfrac{\partial V}{\partial \lambda} - \dfrac{1}{a} \dfrac{\partial U}{\partial \lambda}, \\
D &\equiv \dfrac{1}{a\cos{\varphi}}\dfrac{\partial u}{\partial \lambda} + \dfrac{1}{a\cos{\varphi}}\dfrac{\partial}{\partial \varphi}(u \cos{\varphi}) \notag \\
&= \dfrac{1}{a(1-\mu^2)}\dfrac{\partial U}{\partial \lambda} + \dfrac{1}{a} \dfrac{\partial V}{\partial \mu}.
\end{align}
Temperature is divided into the horizontally and temporally averaged component and the deviations from it, as follows:
\begin{align}
T(\lambda, \varphi, \sigma, t) = \overline{T}(\sigma) + T'(\lambda, \varphi, \sigma, t).
\end{align}
With these new notations, the primitive equations can be re-written as follows.

\paragraph{The continuity equation}

\begin{align}
\dfrac{\partial \pi}{\partial t} + \bm{v}_H \cdot \nabla_\sigma \pi = - D - \dfrac{\partial \dot{\sigma}}{\partial \sigma}.
\end{align}

\paragraph{The hydrostatic equation}

\begin{align}
\dfrac{\partial \Phi}{\partial \sigma} = - \dfrac{R^dT}{\sigma}.
\end{align}

\paragraph{The equation of motion}

\begin{align}
\dfrac{\partial \zeta}{\partial t} &= \dfrac{1}{a}\left( \dfrac{1}{1-\mu^2} \dfrac{\partial V_A}{\partial \lambda} - \dfrac{\partial U_A}{\partial \mu} \right) + \mathcal{D}(\zeta), \label{eq:motion of vor} \\
\dfrac{\partial D}{\partial t} &= \dfrac{1}{a}\left( \dfrac{1}{1-\mu^2} \dfrac{\partial U_A}{\partial \lambda} + \dfrac{\partial V_A}{\partial \mu} \right) - \nabla^2_\sigma (\Phi + R^d \overline{T} \pi + KE ) + \mathcal{D}(D),  \label{eq:motion of div}
\end{align}
where
\begin{align}
U_A(\varphi, \lambda , \sigma) &\equiv (\zeta + f) V- \dot{\sigma} \dfrac{\partial U}{\partial \sigma} - \dfrac{R^dT}{a} \dfrac{\partial \pi}{\partial \lambda} + \mathcal{F}_\lambda , \cos{\varphi}, \label{eq:U_A} \\
V_A(\varphi, \lambda, \sigma) &\equiv -(\zeta + f)U - \dot{\sigma}\dfrac{\partial V}{\partial \sigma} - \dfrac{R^dT'}{a}(1-\mu^2) \dfrac{\partial \pi}{\partial \mu} + \mathcal{F}_\varphi \cos{\varphi} , \label{eq:V_A} \\
KE &\equiv \dfrac{U^2 + V^2}{2(1-\mu^2)}, \\
\nabla^2_\sigma &\equiv \dfrac{1}{a^2(1-\mu^2)}\dfrac{\partial^2}{\partial \lambda^2} + \dfrac{1}{a^2}\dfrac{\partial}{\partial \mu} \left[ (1-\mu^2)\dfrac{\partial}{\partial \mu} \right].
\end{align}


\paragraph{The thermodynamic equation}

\begin{align}
\dfrac{\partial T}{\partial t} = &- \dfrac{1}{a} \left( \dfrac{1}{1-\mu^2} \dfrac{\partial UT'}{\partial \lambda} + \dfrac{\partial VT'}{\partial \mu}  \right) +T'D \notag \\
&- \dot{\sigma} \dfrac{\partial T}{\partial \sigma} + \kappa T \left( \dfrac{\partial \pi}{\partial t} + \bm{v}_H \cdot \nabla_\sigma \pi + \dfrac{\dot{\sigma}}{\sigma} \right) + \dfrac{Q}{C_p} + \mathcal{D}(T) + \mathcal{D}'(\bm{v}). \label{eq:thermodynamic}
\end{align}
.


In \Equref{eq:motion of vor}, \Equref{eq:motion of div}, and \Equref{eq:thermodynamic}, \(\mathcal{D}(\zeta)\), \(\mathcal{D}(D)\) and \(\mathcal{D}(T)\) can be decomposed into the horizontal diffusion, $\mathcal{D}_{HD}$, and the dissipation in the so-called sponge layer, $\mathcal{D}_{SL}$:
\begin{align}
    \mathcal{D}(\zeta) &= \mathcal{D_{HD}}(\zeta) + \mathcal{D_{SL}}(\zeta), \\
    \mathcal{D}(D) &= \mathcal{D_{HD}}(D) + \mathcal{D_{SL}}(D), \\
    \mathcal{D}(T) &= \mathcal{D_{HD}}(T) + \mathcal{D_{SL}}(T). 
\end{align}

\bibliography{main/Ref}{}
\bibliographystyle{aastex/aasjournal}

\end{document}